\documentclass[useAMS,usenatbib]{mn2e}

\usepackage{graphicx}
\usepackage{comment}
\usepackage{color}
\title[Radiative shock oscillation model for the long-term flares of Sgr A*]
 {Radiative shock oscillation model for the long-term flares of Sgr A*}
\author[T. Okuda, C. B. Singh, R. Aktar]
 { Toru Okuda$^{1}$ \thanks{E-mail:bbnbh669@ybb.ne.jp},
  Chandra B. Singh$^{2}$\thanks{E-mail:chandrasingh@ynu.edu.cn},
  Ramiz Aktar$^{3}$\\
  $^{1}$ Hakodate Campus, Hokkaido University of Education, Hachiman-Cho 1-2, Hakodate 040-8567, Japan \\
  $^{2}$  South-Western Institute for Astronomy Research, Yunnan University, University Town,  Chenggong, Kunming 650500, \\ People's Republic of China \\
 $^{3}$ Department of Astronomy, Xiamen University, Xiamen, Fujian 361005, People's Republic of China \\
 }

\usepackage{amsmath}

\begin{document}

\date{Accepted XXX. Received YYY; in original form ZZZ}

\pagerange{\pageref{0}--\pageref{0}} \pubyear{2022}

\maketitle

\label{firstpage}

\begin{abstract}
 We examine time-dependent 2D relativistic radiation MHD flows to develop the shock oscillation model for the long-term flares of Sgr A*. Adopting modified flow parameters in addition to the previous studies, we confirm quasi-periodic flares with periods of $\sim$ 5 and 10 days which are compatible with observations by Chandra, Swift, and XMM-Newton monitoring of Sgr A*. Using a simplified two-temperature model of ions and electrons, we find that the flare due to synchrotron emission lags that of bremsstrahlung emission by 1 -- 2 hours which are qualitatively comparable to the time-lags of 1 -- 5 hours reported in several simultaneous observations of radio and X-ray variability in Sgr A*. The synchrotron emission is confined in a core region of 3 $R_{\rm g}$ size with the strong magnetic field, while the bremsstrahlung emission mainly originates in a distant region of 10 -- 20 $R_{\rm g}$ behind the oscillating shock, where $R_{\rm g}$ is the Schwarzschild radius. The time lag is estimated as the transit time of the acoustic wave  between the above two regions. The time-averaged distribution of radiation shows a strong anisotropic nature along the rotational axis but isotropic distribution in the radial direction. A high-velocity jet with $\sim 0.6c$ along the rotational axis is  intermittently found in a narrow funnel region with a collimation angle $\sim 15^\circ$. The shock oscillating model explains well the flaring rate and the time lag between radio and X-ray emissions for the long-term flares of Sgr A*.

\end{abstract}

\begin{keywords}
 black hole physics--Galaxy: centre--hydrodynamics-- radiation mechanism: thermal -- shock waves.--(magnetohydrodynamics) MHD 

\end{keywords}

\section{Introduction}
Sgr A* is the supermassive black hole found in our galactic centre with a mass of $\sim$ 4 $\times$ 10$^6 M_{\odot}$ and located at 8.27 kpc away (Genzel et al. 2003, Genzel, Eisenhauer \& Gillessen 2010) and exhibits peculiar features of observations. The observed luminosity is five orders of magnitude lower than that predicted by the standard thin disc model (Shakura \& Sunyaev 1973, hereafter, SS73 model)
 and the spectrum of Sgr A* differs from the multi-temperature black body spectra obtained from the  SS73 model. It has led to the emergence of various theoretical models which can explain the spectral properties of such radiatively-inefficient sources. The search led to mainly two theoretical models such as Bondi flow with zero angular momentum (Bondi 1952) and advection dominated accretion flow (ADAF) with high angular momentum (Narayan \& Yi 1994, 1995; Stone, Pringle \& Begelman 1999; Igmenshchev \& Abramowicz 1999, 2000; Yuan, Quataert \& Narayan 2003, 2004). Those theoretical models have been extensively studied (see Narayan \& McClintock 2008; Yuan 2011; Yuan \& Narayan 2014, for review).
The advective accretion flow models could explain well the observations (Das, Becker \& Le 2009; Becker, Das \& Le 2011; Yuan, Wu \& Bu 2012; Li, Ostriker \& Sunyaev 2013).
  Since the pioneering works of magnetized discs with shear instability (Balbus \& Hawley 1991; Hawley \& Balbus 1991), several multidimensional magnetohydrodynamic (MHD) simulation works have been performed.  It has been shown that the inflow-outflow properties of matter around black holes are determined by the magnetic field (Machida, Hayashi \& Matsumoto 2000; Machida, Matsumoto \& Mineshige 2001; Stone \& Pringle 2001; Igumenshchev, Narayan \& Abramowicz 2003; Narayan, Igmenshchev \& Abramowicz 2003; Narayan et al. 2012; Yuan, Bu \& Wu 2012; Yuan et al. 2015). 

  Sgr A* is found in mainly two types of accretion states, namely, flaring and quiescent states, based on multi-wavelength observational studies (Genzel, Eisenhauer \& Gillessen 2010 and references therein). The observations of Sgr A*  show that the flares in X-ray and infrared (IR)  usually last 1 -- 3 hours and occur typically a few times a day. The observed emissions in radio and IR flares vary roughly by $1/2$ and $1-5$ factors (Genzel et al. 2003; Ghez et al. 2004; Eckart et al. 2006; Meyer et al. 2006a, 2006b; Trippe et al. 2007; Yusef-Zadeh et al. 2009, 2011), while the X-ray flare emission varies by more than two orders of magnitude (Ponti et al. 2017). 
 
 Several MHD simulation works have attempted to address the flare phenomena of Sgr A* (Chan et al. 2009; Dexter, Agol \& Fragile 2009; Dodds-Eden et al. 2010; Ball et al. 2016; Ressler et al. 2017). For instance, Ball et al. (2016) demonstrated that the magnetic reconnection process accelerates non-thermal electrons from close to the black holes and  explained the physical process behind the rapid variability of X-ray flares. Ressler et al. (2017) performed general-relativistic (GR) MHD simulations, modeled the emission by thermal electrons and reproduced some of the observed features. Roberts et al. (2017) showed the first fitting of their 2D hydrodynamical simulations to Chandra observations of Sgr A* through Markov chain Monte Carlo sampling, modeling the 2D inflow-outflow solutions.  
An MHD model for episodic mass ejection from regions close to the black holes have been proposed in analogy with solar coronal mass ejections, to explain several features of Sgr A* like light curves and spectra (Yuan et al. 2009; Li, Yuan \& Wang 2017). However,
all of the works mentioned above dealt with high angular momentum flow and tried to address the rapid flares of Sgr A* with a period of  hours.

 Besides the high angular momentum flow like ADAF, low angular momentum flow around the black hole can exhibit the formation of standing shock, which is likely to undergo oscillation or remain stable. The studies of the standing shock in an astrophysical context were pioneered  by Fukue (1987) and then Chakrabarti (1989). Further studies of the low angular momentum flows have been carried out to investigate the flow parameter space responsible for the standing shock formation (Chakrabarti \& Das 2004; Mondal \& Chakrabarti 2006),  2D numerical simulations of the shocks (Molteni, Lanzafame \& Chakrabarti 1994; Molteni, Sponholz \& Chakrabarti 1996; Chakrabarti 1996; Molteni, Ryu \& Chakrabarti 1996; Lanzafame, Molteni \& Chakrabarti 1998), quasi-periodic oscillation (QPO) phenomena (Chakrabarti, Acharyya \& Molteni 2004; Giri et al. 2010; Okuda \& Molteni 2012; Okuda 2014; Okuda \& Das 2015) and effects of cooling, viscosity, and mass outflow on the standing shock (Singh \& Chakrabarti 2011; Kumar \& Chattopadhyay 2013; Aktar, Das \& Nandi 2015; Sarkar \& Das 2016, Aktar et al. 2017). 

 Similarly to other MHD simulation works, 2D time-dependent simulations of the low angular momentum flows  onto black holes  showed that  the magneto-rotational instability (MRI) is very robust in the torus even with a weak magnetic field and that the matter accretes onto the black hole due to the MRI (Proga \& Begelman 2003a, 2003b).
 They show that the intrinsic time variability of the low angular momentum flows can naturally explain some of Sgr A* variability.
 The low angular momentum of the accretion flow around Sgr A* may be attributed to the stellar wind from nearby hot stars orbiting around Sgr A* (Loeb 2004; Mo\'{s}cibrodzka, Das \& Czerny 2006; Czerny \& Mo\'{s}cibrodzka 2008). Motivated by these  works, we  examined the shock oscillating model with low angular momentum for Sgr A* using 2D time-dependent MHD calculations and showed that the magnetized flows yield large modulations of luminosities with a time-scale of $\sim$ 5 and 10 days with an accompanying, more rapid, small modulation with a period of 25 hours (Okuda et al. 2019; Singh, Okuda \& Aktar 2021) which are compatible with luminous flares with a frequency of $\sim$ every half a day, one, five, and ten days in the latest observations by Chandra, Swift, and XMM-Newton monitoring of Sgr A* (Degenaar et al. 2013; Neilson et al. 2013, 2015; Ponti et al. 2015). Recent observations of Sgr A* show  characteristic spectra in radio, near-infrared (NIR), and X-ray bands \citep{key-50,key-6-1,key-59-2}.  These emissions are due to various physical processes such as synchrotron and bremsstrahlung and we need to understand where and how these processes work and how radiation is distributed.
Therefore, the next step in these studies is to examine radiation
MHD accretion flow with low angular momentum. In this paper, we solve the equations of relativistic radiation MHD flows using the special relativistic radiation MHD (RadRMHD) module in the public library software PLUTO, confirm the possible scenario responsible for the long-term flares of Sgr A*, and examine characteristic features of radiation and a time correlation between the synchrotron and bremsstrahlung emissions.

\section{Model equations}
\subsection{Basic equations and magnetic field configurations}
The numerical setup for the present work uses grid-based, finite volume computational fluid dynamics code, PLUTO \citep{key-35, key-32-2}.  
Numerical simulations are carried out by solving the equations of RadRMHD in the quasi-conservative form

\begin{eqnarray}
 \frac{\partial (\rho \gamma)} {\partial t} + \nabla \cdot \left(\rho \gamma \textbf{v}\right) &=& 0,
\end{eqnarray}
\begin{eqnarray}
 \frac{\partial \textbf{m}}{\partial t} + \nabla \cdot [\rho h \gamma^{2} \textbf{vv} - \textbf{BB} - \textbf{EE}] + \nabla p &=& \textbf{G},
\end{eqnarray}
\begin{eqnarray}
 \frac{\partial \varepsilon}{\partial t} + \nabla \cdot [\textbf{m} - \rho \gamma \textbf{v}] &=& G^{0},
\end{eqnarray}
\begin{eqnarray}
  \frac{\partial \textbf{B}} {\partial t} + \nabla \times \textbf{E} &=& 0,
\end{eqnarray}
\begin{eqnarray}
  \frac{\partial E_{r}} {\partial t} + \nabla \cdot \textbf{F} &=& -G^{0},
\end{eqnarray}
\begin{eqnarray}
  \frac {\partial \textbf{F}_{r}}{\partial t} + \nabla \cdot {P}_{r} &=& -\textbf{G}.
\end{eqnarray}
where $\rho$, $\textbf{m}$, $h$, $\textbf{v}$ are the density, the momentum  density, the specific enthalpy and the velocity, respectively. $E_{r}$, $\textbf F$, $P_{r}$ are the radiation energy density, the 
radiation flux, and the pressure tensor as moments of the radiation field, respectively. $G^{0}$ and $\textbf{G}$ are the components of the radiation four-force density given by
\begin{eqnarray}
  \left(G^{0}, \textbf{G}\right)_{\rm comov} = \rho \left[\kappa(E_{\rm r} - a_{\rm R} T^4), (\kappa +\sigma) \textbf{F} \right]_{\rm comv}.
 \end{eqnarray}
 Here, all fields are measured in the fluid's comoving frame, $\kappa$ and $\sigma$ are respectively 
the frequency-averaged absorption and scattering coefficients which are adopted as the Kramer's opacity and 0.4  in our case, and $T$ is the fluid's temperature.  $\gamma$ and $\textbf {B}$ are the Lorentz factor and mean magnetic field, respectively. The electric field in the laboratory frame is given by $\textbf E$ = - $\textbf{v} \times \textbf{B}$. 
Besides, there are quantities 

\begin{eqnarray}
p = p_{\rm g} + \frac{\textbf{E}^{2}+\textbf{B}^{2}}{2},  
\end{eqnarray}
\begin{eqnarray}
\textbf{m} = \rho h \gamma^{2} \textbf{v} + \textbf{E} \times \textbf{B},
\end{eqnarray}
\begin{eqnarray}
\varepsilon = \rho h \gamma^{2} - p_{\rm g} - \rho \gamma + \frac{\textbf{E}^2+\textbf{B}^2} {2},
\end{eqnarray}
which account for the total pressure, momentum density, and energy density of matter and electromagnetic fields, respectively. $p_{\rm g}$ is the gas pressure of the comoving fluid.
A pseudo-Newtonian potential is adopted for representing space-time around non-rotating black hole \citep{key-48} and cylindrical coordinates (R, $\phi$, z) has been used.
 A further closure relation is needed for the radiation fields to relate the pressure tensor $P_r^{ij}$ to $E_r$ and $\textbf{F}_r$, which is given by,
  \begin{eqnarray}
  P_r^{ij}=D^{ij}E_r,
  \end{eqnarray}
  \begin{eqnarray}
  D^{ij}=\frac{1-\xi}{2}\,\delta^{ij}+
 \frac{3\xi-1}{2}\mathbf{n}^i\mathbf{n}^j,
  \end{eqnarray}
  \begin{eqnarray}
  \xi=\frac{3+4f^2}{5+2\sqrt{4-3f^2}},
  \end{eqnarray}
  where $\mathbf{n}=\mathbf{F}_r/\vert\mathbf{F}_r\vert$
  and $f=\vert\mathbf{F}_r\vert/E_r$.
  $\delta^{ij}$ is the Kronecker delta.

To generate magnetic field $\textbf{B}$, we use the vector potential $\textbf{A}$ which is prescribed  
as $\textbf{B} = \nabla \times \textbf{A}$ and consider one simple poloidal magnetic field, same as \citet{key-51-2}, defined by the potential

\begin{eqnarray}
  \textbf{A} = (A_{\rm R}=0,  A_{\rm \phi}= \frac{A_0 z}{rR}, A_{\rm z}=0 ),
\end{eqnarray}
where $r=\sqrt{R^2+z^2}$.
The magnitude of the magnetic field is scaled using the parameter $\beta_{\rm out}
 =8\pi (p_{\rm g})_{\rm out} /B_{\rm out}^2$ which expresses the ratio of gas pressure to
magnetic pressure at the outer radial  boundary $R_{\rm out}$ on the equator, then

\begin{eqnarray}
  A_0=  \rm{sign }(z)  \left(\frac{8\pi (p_{\rm g})_{\rm out}}{\beta_{\rm out}} \right)^{1/2} R_{\rm out}^2,
\end{eqnarray}
where  $(p_{\rm g})_{\rm out}$ and $B_{\rm out}$ are  the gas pressure and the strength of the magnetic field at $R_{\rm out}$. In RadRMHD module of PLUTO, we use an one-temperature model, that is, the electron temperature $T_{\rm e}$ is equal to the ion temperature $T_{\rm i}$, and only the bremsstrahlung emission between ion and electron is taken into account of the cooling source.

\begin{table*}
\centering
\caption{Specific angular momentum $\lambda_{out}$, plasma beta $\beta_{out}$, radial velocity $v_{\rm out}$, the sound velocity $a_{\rm out}$, the input density $\rho_{\rm out}$, the temperature $T_{\rm out}$, the injection height of accretion flow $h_{\rm out}$, the radiation energy density $(E_{\rm r})_{\rm out}$, the input accretion rate ${\dot M}_{\rm input}$ at the outer radial boundary $R_{\rm out}$= 200, where $\rho_{0}$ of $10^{-20}$ g  cm$^{-3}$ is used.}
\vspace{3mm}
\begin{tabular}{@{}cccccccccc} \hline
&&&&&& \\

${\rm model}$ & $\lambda_{out}$ & $\beta_{out}$ & $v_{\rm out}$ & $a_{\rm out}$ & $\rho_{\rm out}$ & $T_{\rm out}$ & $h_{\rm out}$ & $(E_{\rm r})_{\rm out}$ & ${\dot M}_{\rm input}$  \\
  &($R_{g}c$)& & ($c$) & ($c$) & (${\rho}_0$) & ($ \rm K$) & ($R_{\rm out}$) & (${\rho}_0 c^2$) & ($M_{\odot} \rm{yr}^{-1}$)  \\
 \hline

 ${\rm Rad1}$     &   1.3         &      50 & -4.98E-2 & 2.73E-2 & 58.7    & 2.538E9 & 0.4315 & 9.1E-8  & 4.0E-6 \\ 
 ${\rm Rad2}$   &     ''         &      500  & '' & '' & ''   & ''  & '' & '' & " \\
\hline
\end{tabular}
\end{table*}

\subsection{Initial and boundary conditions}
We consider here a mass $M = 4\times 10^6 M_{\odot}$ and a mass accretion rate $\dot M = 4\times 10^{-6} M_{\odot}$ yr$^{-1}$ for Sgr A*.
Hereafter, the coordinates $R$ and $z$ are expressed in the unit of the Schwarzschild radius $R_{\rm g}$ given by $R_{\rm g} =2GM/c^2$ where $G$ and $c$ are the gravitational constant and the light velocity.
 The principle of our oscillating shock model for Sgr A*  is that the hydrodynamical flow with a low angular momentum forms a standing shock in the inner region of the accretion disc if the angular momentum is properly selected and that the shock oscillates in the MHD flow under an appropriate magnetic field.  The method to get the primitive variables of  the density $\rho_{\rm out}$, the angular momentum $\lambda_{\rm out}$, the temperature $T_{\rm out}$,  the radial velocity $v_{\rm out}$, and the disc height $h_{\rm out}$ at the outer R-boundary responsible for such standing shock are shown in \citet{key-46} and \citet{key-56-1}. However, in the present radiation problem, we have to specify another parameter as well, radiation energy density $(E_{\rm r})_{\rm out}$. If the gas is assumed to be optically thin,  $(E_{\rm r})_{\rm out}$ is approximately estimated from $ L_{\rm ff} = 4 \pi R_{\rm out}^2 F_{\rm out}$ and $F_{\rm out} = c \; (E_{\rm r})_{\rm out}$ where $L_{\rm ff}$ and $F_{\rm out}$ are the total volume integral of the free-free emission  and the radiation flux at the outer R-boundary,  respectively. $L_{\rm ff}$ is numerically obtained  from the hydrodynamical simulations without radiation. Using the approximate radiation energy density $(E_{\rm r})_{\rm out}$, we perform further simulations by relativistic radiation hydrodynamic code without magnetic field (RadRHD) and obtain a radiative luminosity $L$ 
 and then a revised $(E_{\rm r})_{\rm out}$. Since the flow is fully optically thin, $L$ must be equal to $L_{\rm ff}$.
 Repeating this process until $L = L_{\rm ff}$ is obtained, we finally get the exact $(E_{\rm r})_{\rm out}$.
The radiation energy density $(E_{\rm r })_{\rm out}$ also gives another condition of the radiation energy density $(E_{\rm r})_{\rm in}$ at the inner boundary radius $R_{\rm in}$ from $ L = 4 \pi R_{\rm in}^2 c (E_{\rm r})_{\rm in}$ assuming an optically thin state also at the inner boundary.

Table 1 gives the primitive variables of the specific angular momentum $\lambda_{out}$, the plasma beta $\beta_{out}$, the radial velocity $v_{\rm out}$, the  sound velocity $a_{\rm out}$, the input density $\rho_{\rm out}$, the temperature $T_{\rm out}$, the injection height of accretion flow $h_{\rm out}$, the radiation energy density $(E_{\rm r})_{\rm out}$, the input accretion rate ${\dot M}_{\rm input}$ at the outer radial boundary $R_{\rm out}$= 200, which are same as those in the previous studies \citep{key-46,key-56-1} except for the angular momentum $\lambda$.
 Here, $\lambda = 1.3$ in unit of  $R_{\rm g}c$ is taken to be a little smaller than the previous  $\lambda = 1.35$, to examine the flow how to behave differently from the previous results. 
As the result, we get 2D hydrodynamical steady flow with a standing shock by 2D radiation hydrodynamical simulations (RadRHD) and the 2D flow  is used as the initial condition for 2D radiation MHD simulations (RadRMHD). Fig.~1 shows the profiles of the density $\rho$, the temperature $T$, the Mach number of the radial velocity, and the radiation energy density $E_{\rm r}$ on the equator in the 2D steady flow, where the standing shock is formed at $R \sim$ 58 in more inward side than $R \sim$ 65 in the previous case of $\lambda$ = 1.35.
As to the parameter of  magnetic field strength, $\beta_{\rm out}$, we select it as 50 and 500 (models Rad1 and Rad2), respectively, from the preliminary simulations. 
 Because  too larger $\beta_{\rm out} ( \gg 500)$ (that is, far weaker strength of the magnetic field) results in a steady state of the flow almost same as non-magnetized hydrodynamical flow.

 The computational domain is $0 \leq R \le 200$ and $-200 \le z \le 200$ with the resolution of $410 \times 820$ cells. 
The adiabatic index for studying the flow has been set as 1.6 for all simulation runs.
In both RadRHD and RadRMHD runs, the same boundary conditions are imposed. At the outer radial boundary, $R_{out}=200$, there are two domains: the disc region where the matter is injected and the atmosphere above the disc region.  The primitive variables in Table 1 are imposed  at the disc region of $R=R_{\rm out}$ and $-h_{\rm out} \le z \le h_{\rm out}$.
The axisymmetric boundary condition is implemented at the inner boundary. At the inner edge of $R_{\rm in}=2$, the absorbing condition is imposed in the computational domain. In the vertical direction, $z=\pm200$,  standard outflow boundary conditions are imposed. In the case of the RadRMHD run, the constant magnetic field is imposed on the outer radial boundary and the strength of the magnetic field $B_{\rm out}$ at the outer radial boundary is
0.35 Gauss for $\beta_{\rm out}$ = 50.

Here, we find the optical thickness across the mesh point $\Delta\tau =\kappa \rho \Delta R$  for the initial conditions (Fig.~1) of MHD flow  given by,
\begin{eqnarray}
  \Delta\tau  \sim 2\times 10^{-28}\left(\frac{\rho}{10^{-16}}\right)^2 \left(\frac{T}{10^9}\right)^{-3.5} \left(\frac{\Delta R}{0.2}\right) \ll 1,
 \end{eqnarray}
 where $\kappa$ is the Kramer's opacity. That means the flow is fully optically thin, including the 
 standing shock.
 Accordingly, there is almost no interaction between matter and radiation and the flow is not influenced by the radiation. However, using the RadRMHD module of PLUTO,
 we can examine the radiation fields, focusing only on the evolution of the 
radiation fields confirmed by some test problems in \citet{key-32-2}.

\begin{table*}
\centering
\caption{Averaged values of the shock location $R_{\rm s}$ on the equator, the total radiative luminosity $L$, the radiative luminosity $L_{\rm Rout}$ from the radial outer boundary, the radiative luminosity $L_{\rm zout}$ from the vertical outer boundary, the total mass outflow rate ${\dot M}_{\rm out}$, the mass inflow rate ${\dot M}_{\rm edge}$ at the inner edge of the flow , the ratio of mass outflow rate ${\dot M}_{\rm Rout}$ from the outer radial boundary to ${\dot M}_{\rm out}$ and the ratio of mass outflow rate ${\dot M}_{\rm jet}$ from the funnel region to ${\dot M}_{\rm out}$, obtained in the simulations.}
\vspace{3mm}
\begin{tabular}{@{}ccccccccc} \hline
&&&&&&& \\
${\rm model}$ & $R_{\rm s}$ & $L$ & $L_{\rm Rout}/L$ & $L_{\rm zout}/L$ & ${\dot M}_{\rm out}$ & ${\dot M}_{\rm edge}$ & ${\dot M}_{\rm Rout}/\dot M_{\rm out}$ & ${\dot M}_{\rm jet}/\dot M_{\rm out} $  \\ 
    &  ($R_{\rm g}$)& (erg $s^{-1}$)& -- &  -- & (${\dot M}_{\rm input}$) & (${\dot M}_{\rm input}$) & -- & -- \\
 \hline
 ${\rm Rad1}$     &126 & 3.16E34 & 0.41 & 0.59 & 0.63   &   0.45   &  0.28 & 0.11 \\
 ${\rm Rad2}$     &114  & 3.47E34 & 0.40 & 0.60 & 0.67  &  0.36  &  0.17 & 0.14 \\ 
 
\hline
\end{tabular}
\end{table*}

\begin{figure}
\begin{center}
\includegraphics[width=0.45\textwidth]{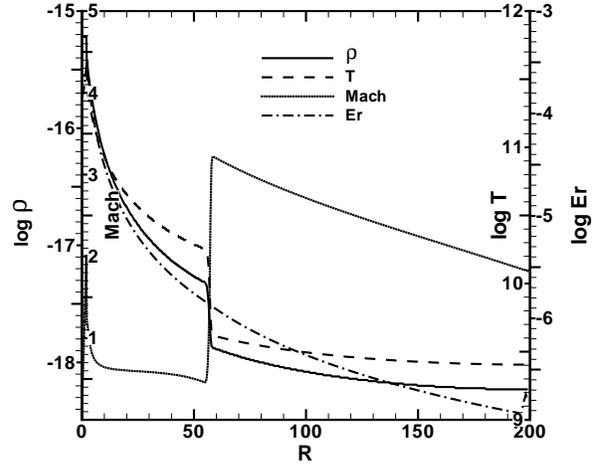}
\end{center}
\caption{ Profiles of the density $\rho$ (g $cm^{-3}$), the temperature $T$ (K), the Mach number of the radial velocity, and the radiation energy density $E_{\rm r}$ (${\rho}_0 c^2)$ on the equator in the steady flow obtained from
2D radiation hydrodynamical simulations. The standing shock is formed at $R \sim 58$.
 }
\end{figure}

\section{Numerical Results}
\subsection{Time variations of luminosity, mass outflow rate and shock location}
The total radiative luminosity $L$ is defined as the sum of $L_{\rm zout}$  and $L_{\rm rout}$ 
which are emitted from the outer z-boundary and the outer R-boundary surfaces, respectively,

 \begin{eqnarray}
  L = L_{\rm zout} + L_{\rm Rout}, 
 \end{eqnarray}
\begin{eqnarray}
  L_{\rm zout} = 2\pi \int_{0}^{R_{\rm out}}\left[ F_{\rm z}(R,Z_{\rm out})-F_{\rm z}(R,-Z_{\rm out})\right]RdR,
\end{eqnarray}
\begin{eqnarray}
 L_{\rm Rout} = 2\pi \int_{-Z_{\rm out}}^{-h_{\rm out}} R_{\rm out} F_{\rm R}(R_{\rm out},z)dz 
 \nonumber \\
   +  2\pi  \int_{h_{\rm out}}^{Z_{\rm out}} R_{\rm out} F_{\rm R}(R_{\rm out},z)dz,
\end{eqnarray}
where $F_{\rm z}(R,z)$ and $F_{\rm R}(R,z)$ are the vertical and radial components of the radiation flux.
The mass outflow rate $\dot M_{\rm out}$ is defined by the total rate of outflow
 through the outer boundaries ($z= \pm Z_{\rm out}$) and   ($R= R_{\rm out}$),

 \begin{align}
 \dot M_{\rm out} &  =  2\pi \int_{0}^{R_{\rm out}} [ \rho (R, Z_{\rm out}) v_{\rm z}(R, Z_{\rm out}) \nonumber \\
 & \;\;\;\;\;\;\;\;\;\;\;-  \rho (R, - Z_{\rm out}) v_{\rm z}(R, - Z_{\rm out}) ] R dR \nonumber \\
  & + 2\pi  \int_{-Z_{\rm out}}^{-h_{\rm out}} R_{\rm out}^2 \rho (R_{\rm out}, z) v_{\rm R}(R_{\rm out}, z) dz  \nonumber \\
 & + 2\pi \int_{h_{\rm out}}^{Z_{\rm out}} R_{\rm out}^2 \rho (R_{\rm out}, z) v_{\rm R}(R_{\rm out}, z) dz
 \end{align}
where  $v_{\rm R}(R,z)$ and $v_{\rm z}(R,z)$ are the radial and vertical components of the velocity.

\begin{figure}
 \begin{center}
 \includegraphics[width=0.45\textwidth]{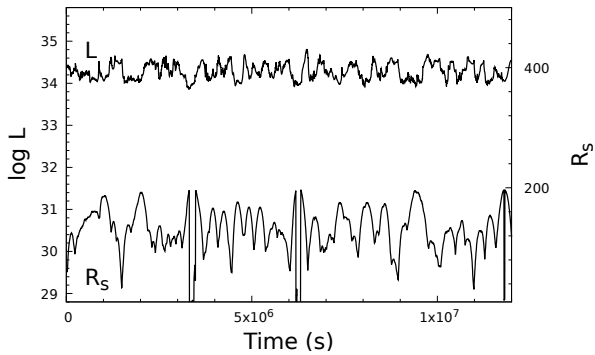}
  \caption {Variation of Luminosity $L$ (erg $s^{-1}$) and shock location $R_{s}$ with time for model Rad1.
  }  
  \end{center}
  \end{figure}

 \begin{figure}
 \begin{center}
  \includegraphics[width=0.45\textwidth]{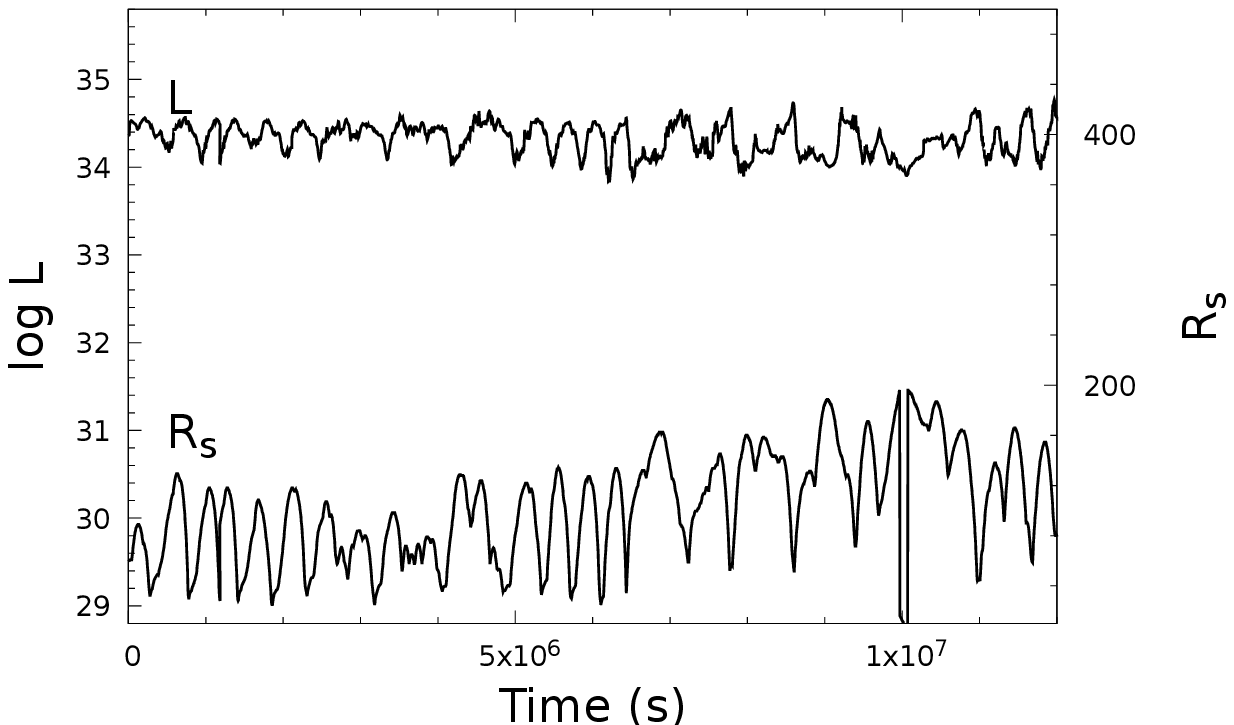}
 \caption {Same as figure 2 but for model Rad2.
  }
 \end{center}
 \end{figure}

  \begin{figure}
\begin{center}
\includegraphics[width=0.45\textwidth]{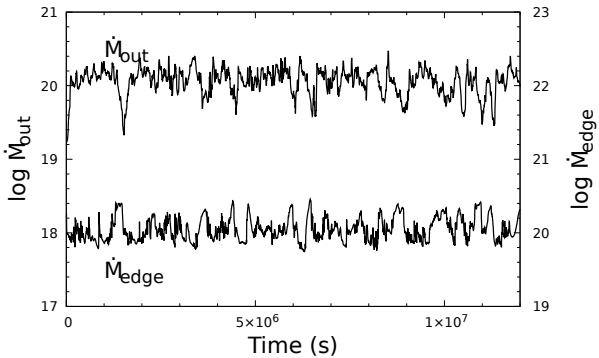}
\caption {Mass inflow $\dot M_{edge}$ (g $s^{-1}$) as well as outflow $\dot M_{out}$ (g $s^{-1}$) rate evolving with time for model Rad1. }
\end{center}
\end{figure}
     
\begin{figure}
 \begin{center}
 \includegraphics[width=0.45\textwidth]{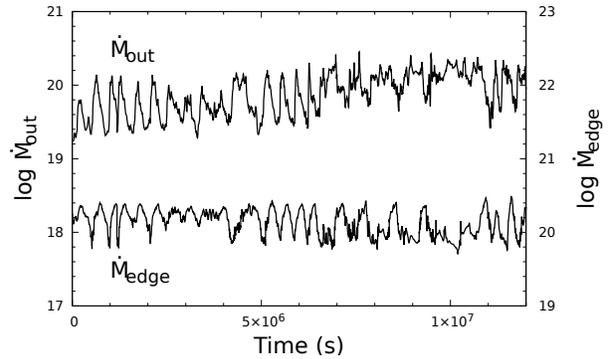}
\caption {Same as figure 4 but for model Rad2. }
 \end{center}
 \end{figure}

Figs.~2 and ~3 show the time evolution of luminosity $L$ and shock location $R_{\rm s}$ on the equator for models Rad1 and Rad2, respectively. In both models, the behaviour of $L$ is opposite to $R_{\rm s}$, that is, there is an increase (decrease) in luminosity when the shock moves towards (away from) the black hole. The luminosity varies around an average value of $\sim 3 \times 10^{34}$ erg  s$^{-1}$ while the shock location varies around an average value of 126 and 114, respectively, for models Rad1 and Rad2.
The corresponding evolution of mass flow rate $\dot M_{\rm edge}$ at the inner edge of the flow and outflow rate  $\dot M_{\rm out}$ are shown in Figs.~4 and ~5 for models Rad1 and Rad2, respectively. Around 40\% of the input gas $\dot M_{\rm input}$ ($\sim 2.4 \times 10^{20}$ g $s^{-1}$) falls onto the event horizon of the black hole and $\dot M_{\rm out}$ is large as $\sim 60\%$ of $\dot M_{\rm input}$. 
Averaged values of the shock location $R_{\rm s}$ on the equator, the total radiative luminosity $L$, the radiative luminosity $L_{\rm Rout}$ from the radial outer boundary, the radiative luminosity $L_{\rm zout}$ from the vertical outer boundary, the total mass outflow rate ${\dot M}_{\rm out}$, the mass flow rate ${\dot M}_{\rm edge}$ at the inner edge of the flow, the ratio of mass outflow rate ${\dot M}_{\rm Rout}$ from the outer radial boundary to ${\dot M}_{\rm out}$, and the ratio of jet mass outflow rate ${\dot M}_{\rm jet}$ to ${\dot M}_{\rm out}$ are listed in Table 2.

\begin{figure*}
    \begin{tabular}{cc}

      \begin{minipage}{0.5\linewidth}
        \centering
        \includegraphics[width=0.8\textwidth]{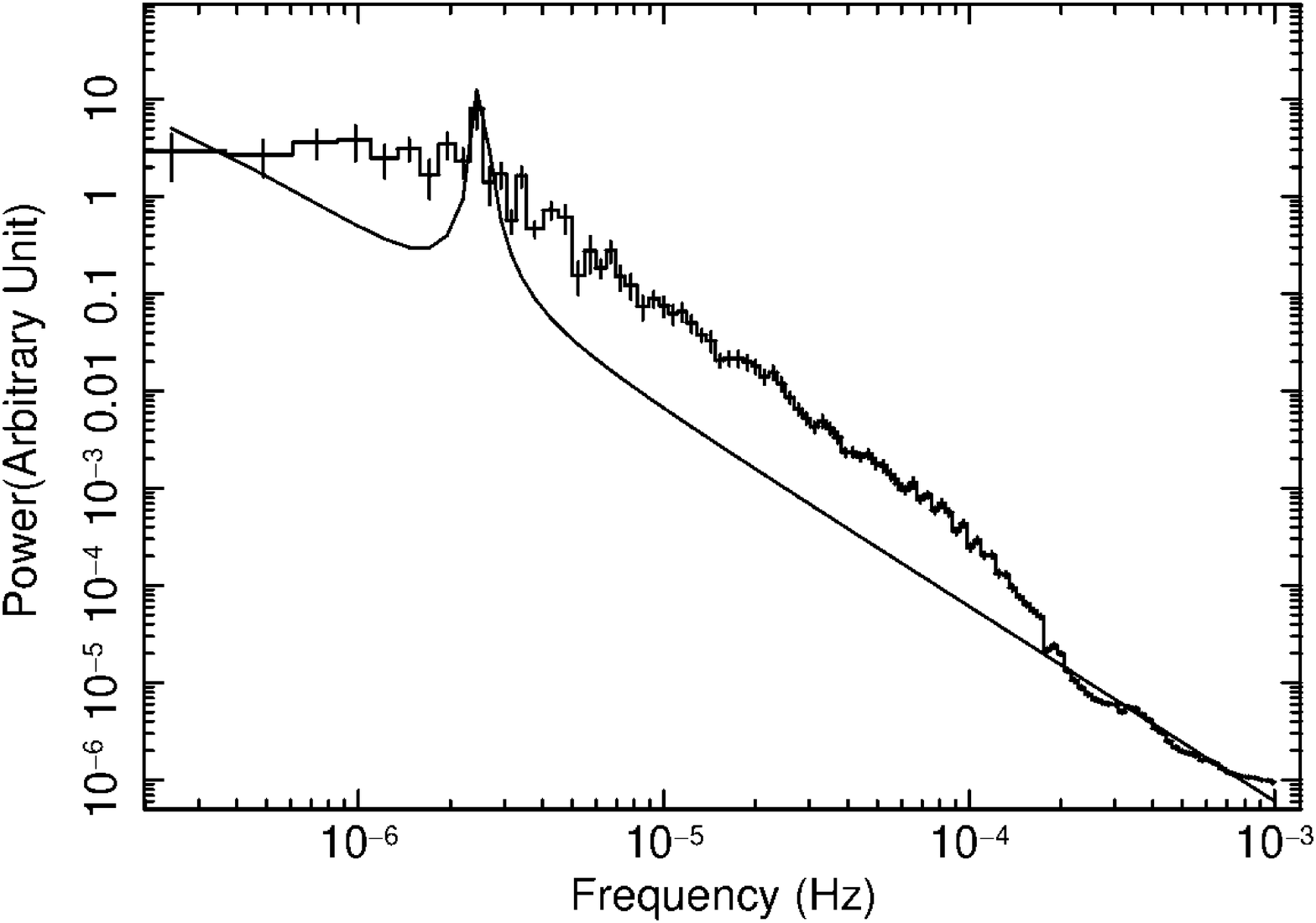}
        \label{fig:3a}
      \end{minipage}

      \begin{minipage}{0.5\linewidth}
        \centering
        \includegraphics[width=0.8\textwidth]{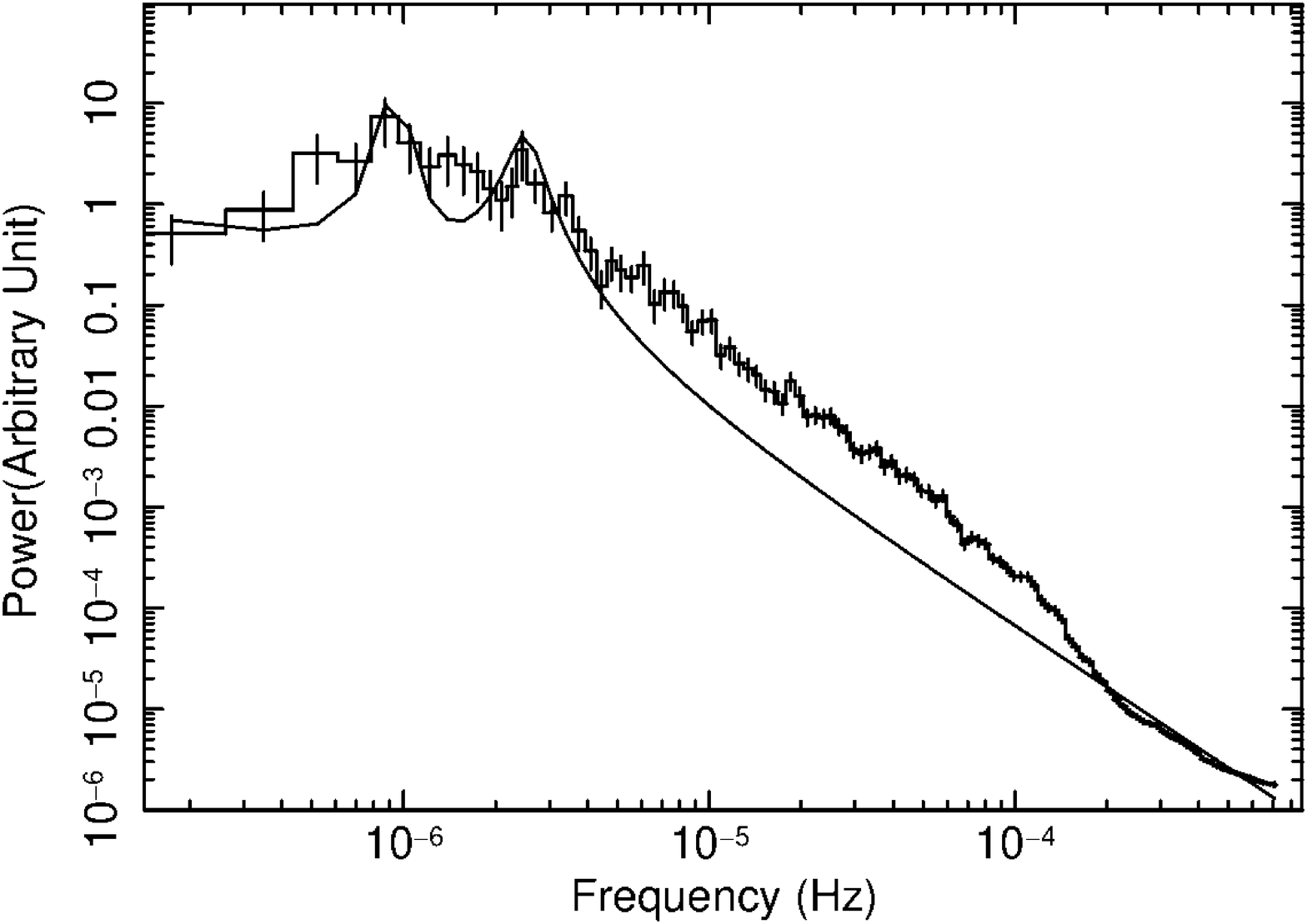}
        \label{fig:3b}
      \end{minipage} \\

    \end{tabular}
\caption {Power density spectra for model Rad1 (left panel) and model Rad2 (right panel).
  }
  \label{Fig: fig6}
  \end{figure*}
  
Fig.~6 show the power density spectra (PDS) of luminosity for models Rad1 (left panel) and  Rad2 (right panel), respectively. For the Rad1 model, the peak (fundamental) frequency is found to be at $2.51 \times 10^{-6}$ Hz which corresponds to $3.98 \times 10^{5}$ s (4.61 days) while in the case of the Rad2 model, there is a primary peak at $9.44 \times 10^{-7}$ Hz corresponding to $1.06 \times 10^{6}$ s (12.26 days) and also a secondary peak at $2.48 \times 10^{-6}$ corresponding to $4.03 \times 10^{5}$ s (4.67 days). Thus, the luminosities are found to vary quasi-periodically with periods of $\sim$ 5 and 12 days.

\subsection{Characteristic magnetized flow with oscillating shock}
The flow is robustly subject to MRI  during the time evolution.
The magnetic field is amplified rapidly by the MRI, and  the MHD turbulence prevails near the equatorial plane. 
Since analyses of the MRI  and the turbulent flow are explained 
in detail in the previous papers (Okuda et al. 2019; Singh, Okuda \& Aktar 2021), we do not repeat them here.
The turbulent flow is spread over $R \sim 100$, and the angular momentum is transported outward.
As the result, the magnetized flow becomes very asymmetric above and below the equatorial plane.
The initial hydrodynamical steady standing shock oscillates quasi-periodically due to the above 
MRI  and turbulence.
The magnetic field intermittently increases near the event horizon, and the magnetic pressure gradient force begins to
dominate the gas pressure gradient force, the gravitational and centrifugal forces along the rotational axis and also in the
equatorial direction. This leads to an intermittent high-velocity jet along the rotational axis and an outflow even in the equatorial
direction. The outflow above and below the equator is at one time faded into the strong accreting flow and grows at other times in the outer turbulent flow as an expanding shock.
The expanding shock sometimes interacts with the contracting oscillating shock and is incorporated into the oscillating shock. 
Thus the outflow driven by the strong magnetic field near the horizon interacts with 
 the accreting gas and the oscillating shock and yields the complicated luminosity variations.

Fig.~7  shows  characteristic features of the magnetized flow at $t= 1.98 \times 10^{7}$ s for model Rad1
 where the thick contours lines through the outer cross ($R\sim 85$) on the equator denote the outer oscillating shock location and the inner thick contour lines through the inner cross ($R\sim 25$) is the expanding inner shock.
 Accreting gas flows strongly into the event horizon hyperbolically across the equator but a part of the gas flows out along the rotational axis and leads to a relativistic jet. In this figure, the jet appears only in the upper region
 of the equator due to the strong asymmetry of the flow. On the other hand, the turbulent flow is dominant  within the expanding inner shock.
Fig.~8 shows the profiles of the density $\rho$, the magnetic field strength $\vert\textbf{B}\vert$, and the Mach number of the radial velocity on the equator at $t= 1.98 \times 10^{7}$ s for model Rad1. The outer oscillating shock
 and the inner expanding shock are found at $R \sim 84$ and 24, respectively. At the evolutionary phase in this figure, the oscillating shock is moving outward just after it undergoes maximum contraction
 and the outflow occurring near the event horizon appears as an expanding shock at $R \sim 24$. 
In the innermost region of $R \le 5$, the density and the strength of the magnetic field are found to be very high.

\begin{figure}
\begin{center}
\includegraphics[width=90mm,height=70mm]{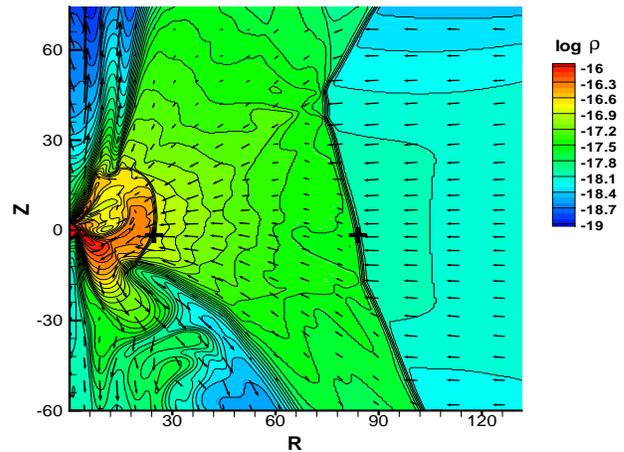}
\end{center}
\caption{ Contours of density $\rho$ (g $cm^{-3}$) with velocity vectors at $t = 1.98 \times 10^7$ s for model Rad1
 Thick contour lines show the outer oscillating shock and the expanding inner shock.
 Crosses show the shock location points on the equator.
 }
\end{figure}

\begin{figure}
\begin{center}
\includegraphics[width=80mm,height=60mm]{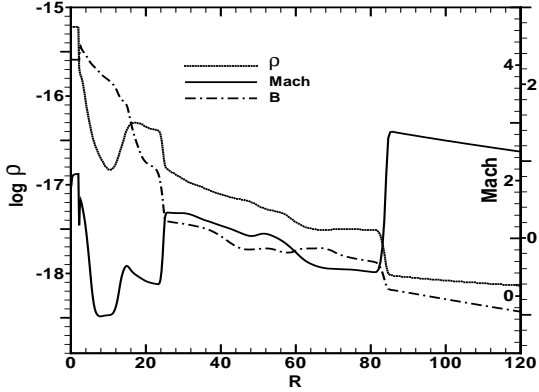}
\end{center}
\caption{  Profiles of density $\rho$ (g $cm^{-3}$), magnetic field strength $\vert\textbf{B}\vert$ (Gauss), and Mach number
 on the equator at $t= 1.98 \times 10^{7}$ s for model Rad1.
 }
\end{figure}

 Fig.~9 shows a schematic diagram for our oscillating shock model. The central compact core $(R \le 5)$ and the turbulent hot region $(5 \le R \le 20)$ behind the expanding inner shock are the dominant  synchrotron and bremsstrahlung emitting regions, respectively, as are mentioned later.

\subsection{Anisotropic radiation}
The numerical results show time-dependent anisotropic radiation and asymmetric natures of
 flow. The observed radiation comes through the outer z-boundary and R-boundary above or below the equatorial plane.
 Let  $\theta$ be the angle at the center measured from the rotating axis.
 The luminosity emitted per unit solid angle  ${\rm d }{\rm  \Omega}$ at angle $\theta$ is given by 

 \begin{equation}
 \frac{{\rm dL} } {{\rm d }{\rm  \Omega}} = ( F_{\rm z} \; {\rm cos} \theta + F_{\rm R} \;{\rm sin} \theta ) \times (R^2 + z^2),
\end{equation}
where $F_{\rm R}$ and $F_{\rm z}$ are the R and z components of radiation flux $\textbf{F}$, and ($R$, $z$) is the 
coordinates of a point on the outer $R$ and $z$  boundary surfaces
through which the light from the central core passes.

\begin{figure}
\begin{center}
\includegraphics[width=60mm,height=80mm, angle=90]{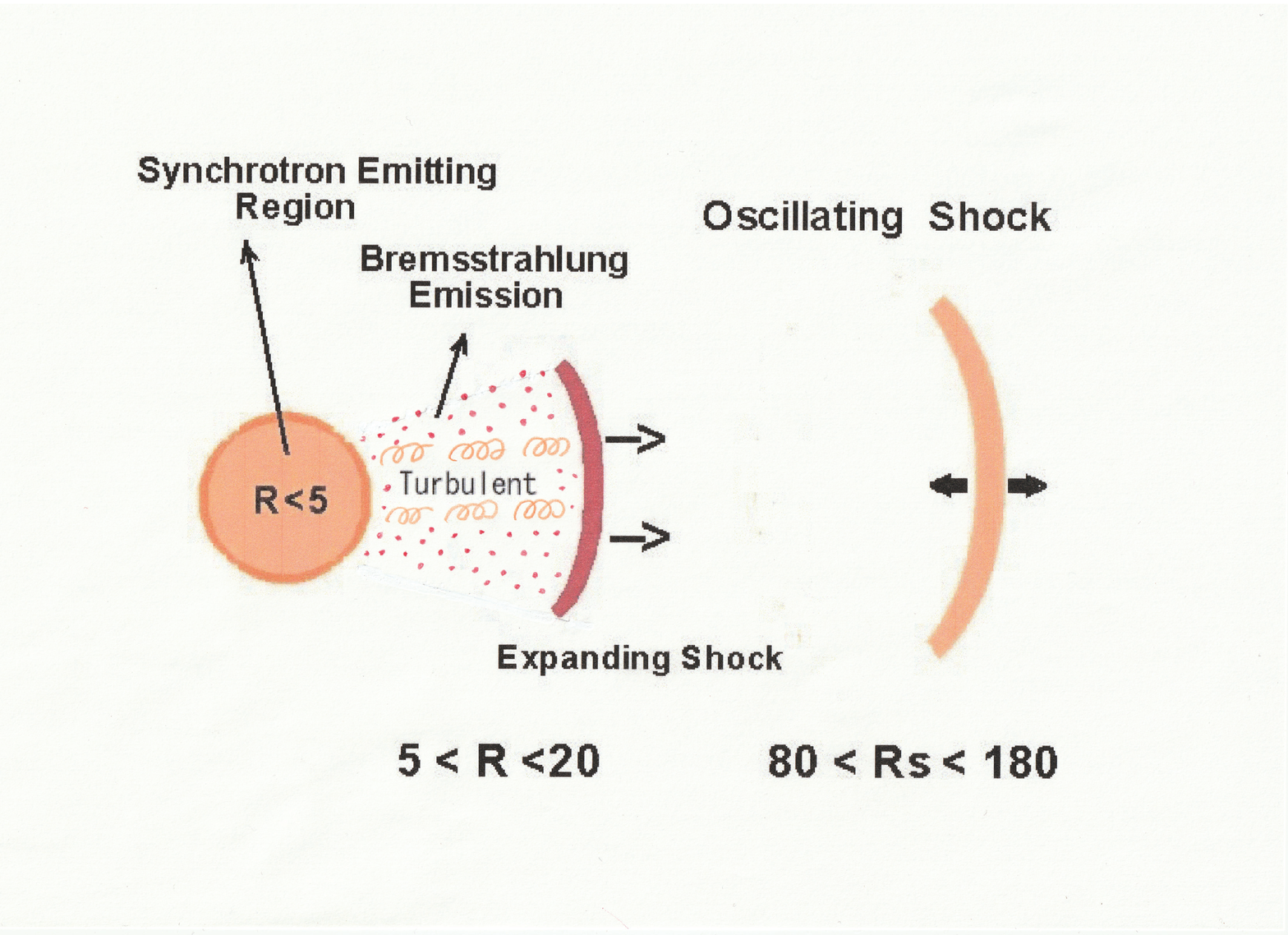}
\end{center}
\caption{ Schematic diagram of the oscillating shock model. 
 Owing to the MRI, outer standing shock oscillates quasi-periodically and the flow becomes turbulent in the central region. The sporadically increasing magnetic field near the event 
horizon yields to an intermittent outflow  and the outflow grows as an expanding shock.  
When the oscillating shock undergoes maximum contraction and interacts with the expanding shock, the maximum bremsstrahlung luminosity is attained. To the contrary, when the oscillating shock moves far away, the luminosity becomes minimum.
 }
\end{figure}

\begin{figure}
\begin{center}
\includegraphics[width=80mm,height=110mm]{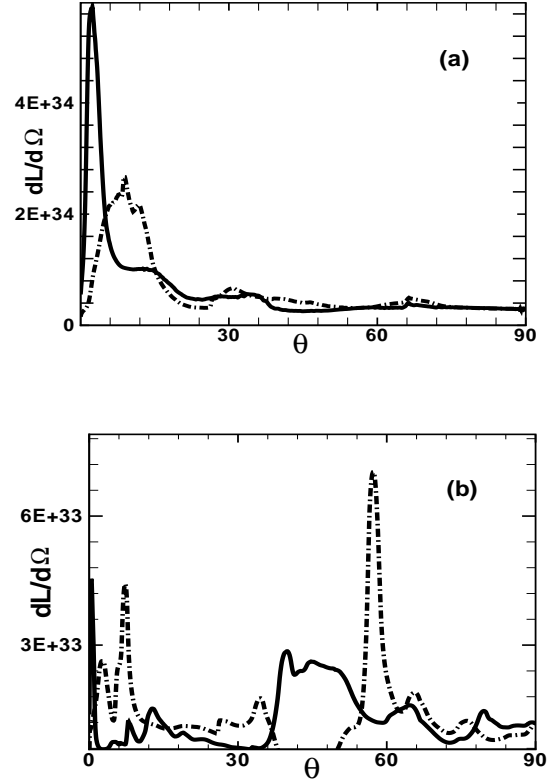}
\caption{Anisotropic distributions of radiation ${\rm dL}/{\rm d} \Omega$ (erg s$^{-1}$)
 per unit solid angle at angle $\theta$ measured from the rotational axis
 above (solid line) and below (dash-dot line)  the equatorial plane at two different times of  (a)  $2.0 \times 10^7$ and  (b) $ 2.4 \times 10^7 $ s for model Rad1.
 }
\end{center}
\end{figure}

\begin{figure}
\begin{center}
\includegraphics[width=80mm,height=60mm]{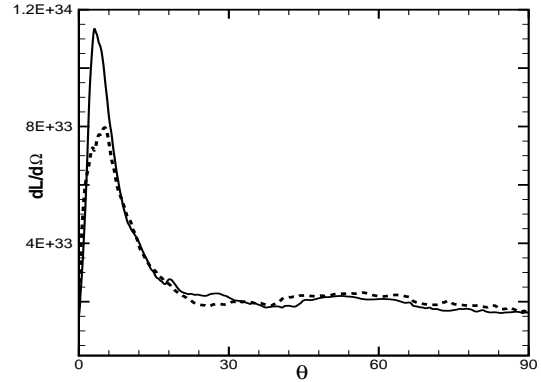}
\caption{Same as fig.~10  but averaged one over a time duration of  $ 9.6 \times 10^6$ -- $2.4 \times 10^7$ s for model Rad1.
The radiation distribution shows an anisotropic property along the rotational axis on the outer z-boundary ( $0 \le \theta \le 45^\circ$ ) but isotropic nature 
  on the outer R-boundary ($45^\circ \le \theta \le 90^\circ$).
 }
\end{center}
\end{figure}

 Fig.~10 shows the radiation distribution ${\rm dL}/{\rm d} \Omega$ per unit solid angle
at  angel $\theta$ measured from the rotational axis
 above (solid line) and below (dash-dot line) the equatorial plane at two different
 times of   $2.0  \times 10^7$ and $2.4 \times 10^7 $ s 
 for model Rad1. The radiation distribution in (a) is concentrated along
 the rotational axis but in (b) it varies over a wide region.
 Generally, the radiation distributions are different depending on time
  and  above and below the equatorial plane. However, the averaged distribution of radiation 
  over a long time (Fig.~11) shows an anisotropic distribution along the rotational axis both above and below the equatorial plane.
  The strength of the isotropic radiation through the outer radial boundary
 $ (45^\circ \le \theta \le 90^\circ )$ is one-sixth to a quarter of the maximum strength near the rotational axis. The luminosity $L_{\rm Rout}$ emitted from the outer radial boundary amounts to two-thirds of the luminosity $L_{\rm zout}$ from the outer z-boundary surface 
 because the area in the former region is larger than the latter region.

\subsection{Mass outflow and high velocity jet}
 As is mentioned in the subsection 3.1, most half of the accreting gas is swallowed into  the black hole
 and the remainder  flows out along the rotational axis and in the turbulent  region 
 above and below the equatorial plane.  The intermittently  increasing magnetic field near the horizon  plays an important role in these outflows.
  The outflow along the rotational axis is accelerated by the magnetic pressure gradient force 
   and develops as a high-velocity jet.  
   The magnetic-pressure gradient force dominates  the gravitational force and the
 gas-pressure gradient force in the upper region $R > 50$ within a funnel region along the rotational axis.
   Fig.~12 shows a high-velocity jet phenomenon at $t= 1.98 \times 10^{7}$ s for model Rad1.
 The jet is formed at a collimation angle $\sim 15^\circ$ within the funnel region and the jet velocity amounts to $\sim 0.6 c$ at the outer surface.
 Compared with the results in the previous case with  $\lambda$=1.35 \citep{key-46},
 the jet is strongly accelerated and collimated because the magnetic field  used 
 is taken to be larger by one order of magnitude.
 Differently from the strong jet phenomenon in the upper region above the equator, the high-velocity jet is not found below the equator at $t= 1.98 \times 10^{7}$ s.

\begin{figure}
\begin{center}
\includegraphics[width=100mm,height=90mm]{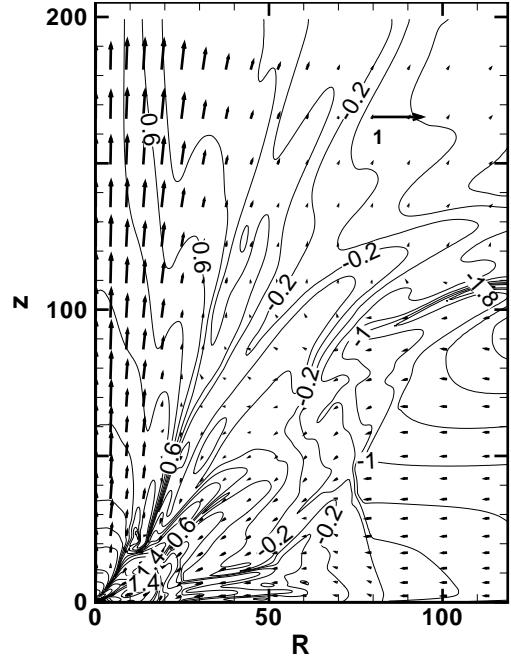}
\end{center}
\caption{Velocity vectors with contours of the magnetic field at  $ 1.98 \times 10^7$ s 
 for model Rad1. The contour labels denote the strength log $B$ (Gauss) of magnetic field.
 The jet at the outer surface attains to the velocity $\sim 0.6$ and is collimated in a narrow angle $\theta \sim 15^\circ$.
 }
\end{figure}.
 
 Despite asymmetric features of the flow to the equatorial plane, the averaged mass outflow rates from the outer z-boundary surfaces are $\sim 1.8 \times 10^{-6} M_{\odot}\:{\rm yr}^{-1}$ (60$\%$ of the input accretion rate). 
 The averaged mass-outflow rate of the jet  is $\sim 10 \%$ of the total mass outflow rate in models Rad1 and Rad2.
 However, we notice that these jet phenomena intermittently occur. The maximum outflow rate occurs generally at the phase of maximum shock location but the high-velocity jet
 appears at the phase of minimum shock location, that is, maximum luminosity.
 
\subsection{Synchrotron and bremsstrahlung emissions}
To examine the contribution of synchrotron and bremsstrahlung emissions to the total luminosity,
we adopt a simplified two-temperature model of the ion temperature $T_{\rm i}$ and the electron temperature $T_{\rm e}$, assuming
that the ratio $\alpha= T_{\rm e} /T_{\rm i}$ is constant in the region considered here.
Then, the bremsstrahlung cooling rate $q_{\rm br}$ is given as follows (Stepney \& Guilbert 1983).
   \begin{eqnarray}
  q_{\rm br} = q_{\rm ei} + q_{\rm ee},
  \end{eqnarray}
  
  \begin{eqnarray}
  q_{\rm ei}=1.48\times 10^{-22}n_{\rm e}^2 F_{\rm ei}(\theta_{\rm e}) \hspace{1cm}{\rm erg \ cm^{-3}\; s^{-1}},
  \end{eqnarray}
  
  \begin{eqnarray}
   \;\;\;F_{\rm ei}(\theta_{\rm e}) =\left\{\begin{array}{ll}
    1.02\theta_{\rm e}^{1/2}(1+1.78\theta_{\rm e}^{1.34}) & \mbox{ for $\theta_{\rm e} <1$},  \\
    1.43\theta_{\rm e}[{\rm ln}(1.12\theta_{\rm e}+0.48)+1.5]
                                                          & \mbox{ for  $\theta_{\rm e}>1$},
                                              \end{array}
                                      \right. 
  \end{eqnarray}
  
  \begin{eqnarray}
  q_{\rm ee} = \left\{\begin{array}{llll}
     2.56\times 10^{-22}n_{\rm e}^2\theta_{\rm e}^{1.5}
       (1+1.10\theta_{\rm e}+\theta_{\rm e}^2-1.25\theta_{\rm e}^{2.5})  & \\
     \hspace{3cm}  {\rm erg \ cm^{-3} \; s^{-1} }  \mbox{ for $\theta_{\rm e} <1$}, 
  & \\
    3.40\times 10^{-22}n_{\rm e}^2\theta_{\rm e} [{\rm ln}(1.123\theta_{\rm e})+1.28 ]
  & \\                                                                                       
     \hspace{3cm}   {\rm erg \ cm^{-3} \; s^{-1}}  \mbox{ for  $\theta_{\rm e} >1$}, & \\
                               \end{array}
                      \right.
    \end{eqnarray}
  where $n_{\rm e}$ and $n_{\rm i}$ are the number density of electrons and ions, $K_0$, $K_1$
  and $K_2$ are
 modified Bessel functions, and the dimensionless electron and ion temperature are defined by

  \begin{eqnarray}
  \theta_{\rm e}= {kT_{\rm e}\over {m_{\rm e}c^2}}, \;\;\; \theta_{\rm i}={kT_{\rm i}\over {m_{\rm p}c^2}}.
  \end{eqnarray}

 The synchrotron cooling rate  $q_{\rm syn}$ is given by (Narayan \& Yi 1995; Esin et al. 1996)

  \begin{eqnarray}
  q_{\rm syn}  &=& 
    {{2{\rm \pi} kT_{\rm e}\nu_{\rm c}^3 }\over {3Hc^2}}  
     +6.76\times 10^{-28} {n_{\rm i}\over{K_2(1/\theta_{\rm e})a_1^{1/6}}} \nonumber \\ 
  & \times &  [{1\over a_4^{11/2}}\Gamma({11\over 2},a_4\nu_{\rm c}^{1/3})
     +{a_2\over a_4^{19/4}} \Gamma({19\over 4},a_4\nu_{\rm c}^{1/3}) \nonumber \\        
  &+& {a_3\over a_4^4}( a_4^3\nu_{\rm c}+3a_4^2\nu_{\rm c}^{2/3}    
   +6a_4\nu_{\rm c}^{1/3}+6) {\rm e}^{-a_4\nu_{\rm c}^{1/3}}] \nonumber \\   
  && \hspace{3.8cm} {\rm erg \ cm^{-3} \ s^{-1}},
   \end{eqnarray}
   where  $H$ is a height of the disc,
    \begin{eqnarray}
   a_1={2\over {3\nu_0\theta_{\rm e}^2}},\; a_2={0.4\over a_1^{1/4}},\; a_3={0.5316\over a_1^{1/2}},
   \; a_4={1.8899a_1^{1/3}}, \nonumber \\                                    
   \Gamma(a,x)= \int_{x}^{\infty}t^{a-1}{\rm e}^{-t}dt, 
   \nu_0={eB\over {2{\rm \pi} m_{\rm e}c}} {\; \rm and} \;\nu_{\rm c}={3\over 2}\nu_0\theta_{\rm e}^2x_{\rm M}.
     \end{eqnarray}
     Here $B$ is the strength of the magnetic field and $x_{\rm M}$  is determined from the next equation
     as
    \begin{eqnarray}
      {\rm exp}(1.8899x_{\rm M}^{1/3} ) &=&  2.49\times 10^{-10}{4{\rm \pi} n_{\rm e}r\over
       B} {1\over {\theta_{\rm e}^3 K_2(1/\theta_e) }} \nonumber \\
       &\times&    \left({1\over  x_{\rm M}^{7/6} }
        + {0.40 \over x_{\rm M}^{17/12}}+{0.5316\over {x_{\rm M}^{5/3}}}\right). \nonumber \\        
    \end{eqnarray}
 
  Since some two-temperature advection-dominated accretion models for black holes show that $\alpha\sim$  0.01 -- 0.05	 \citep{key-42} and 0.2 -- 0.06 \citep{key-32-1} in the inner region of
 $ 10 \le R \le 100 $, we assume two cases of $\alpha$ = 0.01 and 0.05 here.  Then, using the primitive variables in the simulations, 
we estimate the luminosities $L_{\rm syn}$ and $L_{\rm brem}$ by the synchrotron and bremsstrahlung emissions, respectively, as follows,

\begin{eqnarray}
  L_{\rm syn} =\int q_{\rm syn} {\rm d} V,
\end{eqnarray}

\begin{eqnarray}
  L_{\rm brem} =\int q_{\rm brem} {\rm d} V,
\end{eqnarray}
where the volume integration is done over all computational zones.

The time-variations of $L_{\rm syn}$ and $L_{\rm brem}$ for  $\alpha$ = 0.01 and 0.05 are shown in Figs.~13 and ~14, respectively.
In both cases of $\alpha$ = 0.01 and 0.05, the synchrotron luminosity is more
 than one order of magnitude larger than the bremsstrahlung one.
Furthermore, $L_{\rm syn}$ and $L_{\rm brem}$ for $\alpha=0.05$ are far higher than those for $\alpha$=0.01 because the electron temperature in the former is high, and the amplitude of  luminosity variations in the former are large.
These results show that the relevant radiative luminosities given in Table 2 would be much larger if we take account of the two-temperature model in PLUTO code
although the flow structures are not largely altered because the gas is fully optically thin.

\begin{figure}
\begin{center}
        \includegraphics[width=80mm,height=50mm]{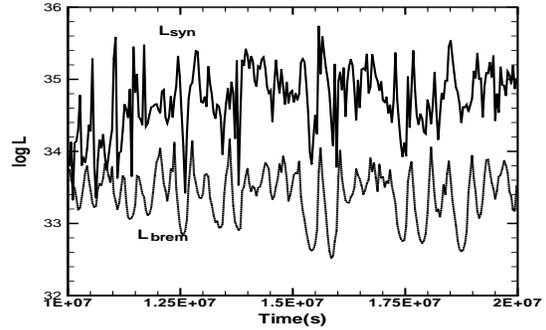}
\caption {Time variations of luminosities $L_{\rm syn}$ and $L_{\rm brem}$ (erg s$^{-1}$) for $\alpha=0.01$ for model Rad1.
  }
\end{center}
\end{figure}

 \begin{figure}
\begin{center}
        \includegraphics[width=80mm,height=50mm]{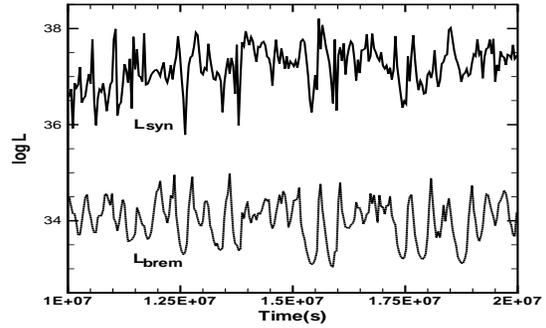}
\caption {Same as figure 13 but for 0.05 for model Rad1.
  }
 \end{center}
 \end{figure}

\subsection{Time-correlation between synchrotron and bremsstrahlung emissions}
 If the synchrotron and bremsstrahlung emissions are emitted in different regions or occur originally with a time delay, we may observe any time lag between these emissions.
 To examine the time correlation between both emissions, in Fig.~15, we plot
 the luminosity variations of  $L$, $L_{\rm syn}$, and $L_{\rm brem}$ every time interval  of 100$R_{\rm g}/c \; ( 3960$ s $\sim$ one hour)  during  $t$ = 2.4$\times 10^6$ -- $2.8\times 10^6$ s for model Rad1. The time interval  100$R_{\rm g}/c $ is the maximum time resolution for the output of the primitive variables in our simulations.

\begin{figure}
\begin{center}
\includegraphics[width=90mm,height=70mm]{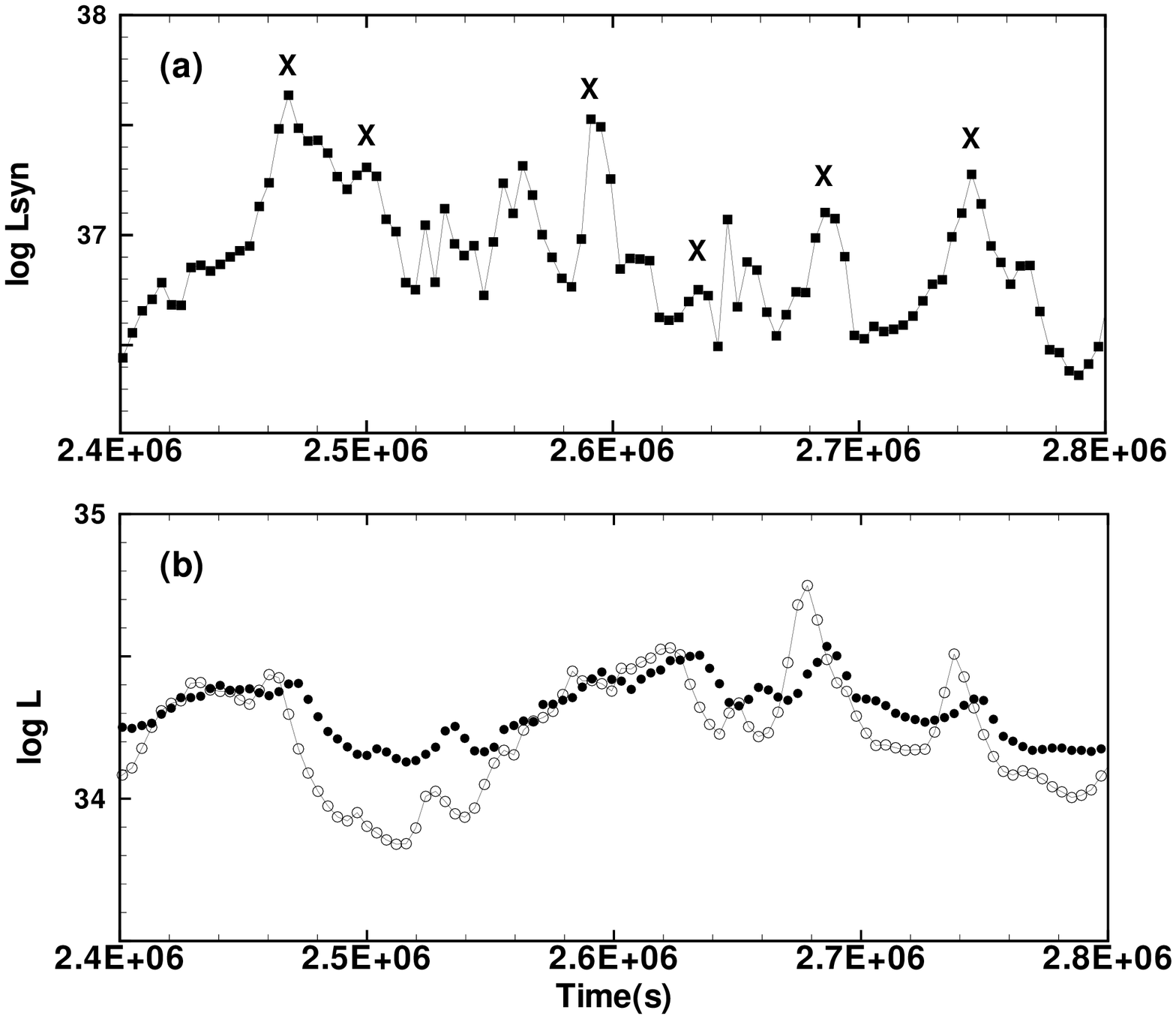}
\end{center}
\caption{Time variations every time interval $100R_{\rm g}/c$ of luminosities  $L_{\rm syn}$, $L_{\rm brem}$, $L$  (erg s$^{-1}$) for $\alpha (=T_|{\rm e}/T_{\rm i}) $=0.05 of model Rad1.
(a) : $L_{\rm syn}$ (filled square) and (b) : $L_{\rm brem}$
 (open circle) and $L$(filled circle).  Crosses in (a) present the local maximum peaks of $L_{\rm syn}$. 
 }
\end{figure}

From the positions of local maximum peaks in the lower panel (b), we find the maximum  $L$ emitted from the outer boundary surface lags that of $L_{\rm brem}$ roughly by 2 --3 points, that is, 2 -- 3 hours. 
Since the effective bremsstrahlung emitting region is mainly near the equatorial plane, the time-lag is reasonably explained in terms of the light crossing time of $200 R_{\rm g}/c$ ($\sim$ 2 hours) of the equatorial plane to the outer z-boundary  surface. 
On the other hand, from a comparison of the local maximum peaks of $L_{\rm brem}$ and $L_{\rm syn}$, the synchrotron radiation lags the bremsstrahlung emission by  1 --  2 points interval 
(1 -- 2 hours).  The time lag is not regarded  as the light crossing time between the different emitting regions because the distance between the different regions is considered to be far smaller than that between the equator and the outer z-boundary surface. However, it is conceivable that the flare due to the synchrotron radiation originally occurs in a time delay after the bremsstrahlung  flare, if the flares occur due to the same thermal process of shock heating and each emitting region is separated to some extent. As is mentioned in subsection 3.2,  the bremsstrahlung luminosity becomes maximum when the oscillating shock contracts mostly, and the post-shock temperature and density are enhanced considerably. Therefore, the maximum synchrotron emission may occur at a delay time after suffering a strong perturbation of the thermal process same as the bremsstrahlung, if the synchrotron emission originates in a far inward region, compared with the bremsstrahlung emission region. 
The delay time is $\sim$ a transit time of the acoustic wave between two emitting regions.
 
 To examine the effective emitting regions of $L_{\rm syn}$ and $L_{\rm brem}$, we calculate local synchrotron and bremsstrahlung luminosities $L_{\rm syn}(R)$  and $L_{\rm brem}(R)$ which are emitted within a sphere of radius $R$. Fig.~16 shows the time variations for the ratio of the local luminosities $L_{\rm syn}(3)$ (open diamond),  $L_{\rm syn}(5)$ (filled circle), $L_{\rm brem}(5)$ (open circle), and $L_{\rm brem}(10)$ (solid line)
 to their total luminosities, respectively, during time of 4 -- 8$\times 10^7$ s,
  where, for $L_{\rm syn}(3)$, only the values corresponding to the local maximum luminosity are plotted because of its visibility of the plot. We find here that most of the synchrotron emission is emitted within a compact region of $R \le 3$ but the bremsstrahlung emission comes mostly from a distant region of $ 10 \le R \le 20$ and the contribution of the bremsstrahlung emission from the inner region $R \le 5$ is negligible.
 This is due to that the strength of the magnetic field is mostly strong within the compact small region, while, in the outer region of $10 \le R \le 20$, 
 the magnetic field strength is weak and the density and the temperature are high over a broad region behind the expanding inner shock, as is found in Figs.~7 and ~8. Therefore, the perturbed waves in the bremsstrahlung emitting region at the maximum $L_{\rm brem}$ phase attain to the inward synchrotron emitting region after a transit time $\sim$ (10 -- 20) $R_{\rm g}/c_{\rm s}$ of the acoustic wave where $c_{\rm s}$ is the sound speed. 
 In the inner region $10 \le R \le 20$ at the maximum bremsstrahlung luminosity phase,  $c_{\rm s}$ is $\sim$ 0.1$c$ in our models. Then, the transit time is (100 --200) $R_{\rm g}/c  \sim$ (1 -- 2) hours.
The time lags of 1 -- 2 hours between the maximum peaks of $L_{\rm syn}(R)$  and $L_{\rm brem}(R)$ found in Fig.~15 are well compatible with the transit time of the sound wave between the two different emitting regions.

\begin{figure}
\begin{center}
\includegraphics[width=90mm,height=70mm]{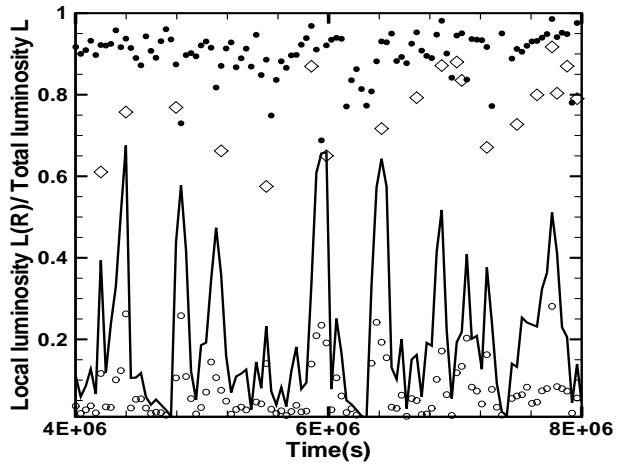}
\end{center}
\caption{Ratio of local synchrotron and bremsstrahlung luminosities $L_{\rm syn}(3)$ (open diamond),  $L_{\rm syn}(5)$ (filled circle), $L_{\rm brem}(5)$ (open circle), and $L_{\rm brem}(10)$ (solid line) to their total ones during time of 4 -- 8$\times 10^6$ s,
 where $L_{\rm syn}(R)$  and $L_{\rm brem}(R)$ are the synchrotron and bremsstrahlung luminosities, respectively, emitted within a sphere of
  radius $R$.
 For $L_{\rm syn}(3)$, only the values corresponding to the local maximum luminosity are plotted because of its visibility of the plot. 
 We find here that most of the synchrotron emission is emitted within a compact region of $R \le 3$ but the bremsstrahlung emission comes mostly from a distant region of $ 10 \le R \le 20$ and the contribution from the inner region $R \le 5$ is negligible.
}
\end{figure}

\section{Observational Relevance to Sgr A*}
The long term  flares over days of Sgr A* has been detected  from radio, sub-millimetre, and IR to X-ray. 
The Chandra X-ray observations of 2006 Feb. to 2011 Oct. and 2012
 show that the flares with X-ray luminosity $L > 10^{34}$ erg s$^{-1}$ occur at a rate of $\sim$ 1.1
 per  day, while luminous flares with $L > 10^{35}$ erg s$^{-1}$  at 0.1 and  0.2  per day
  (Degenaar et al. 2013; Neilsen et al. 2013, 2015; Ponti et al. 2015).
 In XMM-Newton and Chandra monitoring of Sgr A* over fifteen years, 
 the 2012 Chandra observations detect weak to brighter flares occurring at a rate of 1.1
 per day but very bright flares at 0.26 per day, and the synthetic observations with XMM-Newton, Chandra, and Swift shows the flaring rates of 0.27 and 2.5 per day (Ponti et al. 2015).
 These observations constrain the long-term flares to occur approximately 
 every  half a day, one,  five, and ten days. On the other hand,  the flaring rates are
 suggested to change on time scales of years \citep{ key-1-1}.
  In our magnetized flows, the luminosity varies as 
 $\sim 10^{33}$ -- $10^{35}$ erg s$^{-1}$ and the PDS analyses of the luminosity variations confirm the long-term flares to occur every 1, 5, and 10 days, including the previous results.  These flaring rates are well compatible with the above observational flaring rates of Sgr A* which are not yet established statistically but are definitely confirmed in many flares.

\begin{figure}
\begin{center}
\includegraphics[width=90mm,height=60mm]{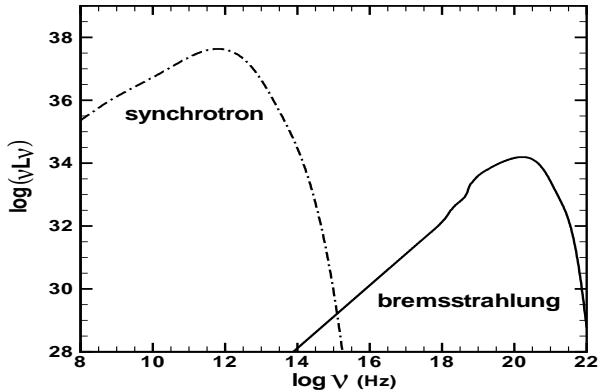}
\end{center}
\caption{Energy spectra $\nu L_{\rm \nu}$ of bremsstrahlung (solid line) and synchrotron (dashed-dot line) radiation$\alpha$=0.05 at $1.98\times 10^7$ s for  model Rad1.
 }
\end{figure}

We showed that the synchrotron emission lags the bremsstrahlung emission roughly by $\sim$ 1 -- 2 hours. Referring to \citet{key-32-1}, we calculate the energy spectra in Fig.~17 and confirm that the synchrotron and bremsstrahlung emissions peak in $\nu \sim
 10^{12}$ and $10^{20}$ Hz, respectively, that is, in radio and X-ray bands. Accordingly we expect a time lag of $\sim$ 1 -- 2 hours between radio and X-ray observed in Sgr A*. The time-lag between radio, near-infrared (NIR), and X-ray in Sgr A* have been reported in several observations. For instance, Yusef-Zadeh et al. (2006, 2008) show time-lags of the flares between multi-wavelength bands, such as the 20 -- 40 min between 22 and 43 GHz and the $\sim$ 2 hour's lag between submillimeter and X-ray in Sgr A*, Yusef-Zadeh et al. (2009) report an X-ray/IR flare with a radio flare delayed by  $\sim 5$ hours but the radio flare delayed by several hours after the X-ray flare, Rauch et al. (2016) detect an NIR flare followed by a radio flare 4.5 hours later,  Capellupo et al. (2017)  find significant radio peak $\sim $ 3 hours later after the brightest X-ray flare and also other associated X-ray and radio variability, with the radio peak appearing $\sim $ 80 minutes later after  weaker X-ray flares, and Witzel et al. (2021)  report sub-millimeter variations which lag those in the NIR by $\sim$ 30 minutes. However, at present, clear statistical evidence of the time-lag relation between X-ray flares and radio variability  has not been  still established. 
 Nevertheless, the time-lags between radio and X-ray flares in our models agree qualitatively with the above observations.
Future simultaneous, long-duration X-ray-radio monitoring of Sgr A* may confirm the time-correlation between X-ray, NIR, and radio.

 In spite of a lot of observations of Sgr A* flares,
the powerful jets of Sgr A* have not been convincingly reported, although an outflow from the
 accretion flow onto Sgr A* is suggested from the extremely weak H-like Fe K$_{\rm \alpha}$ \citep {key-59-1}.
 As is suggested in subsection 3.4,  the intermittent high-velocity jet may be observed although it is difficult to detect the jet in the optical and UV bands due to the distinction of dust through the Galactic plane.
 Perhaps, the detection of the high-velocity jet may be relevant to the inclination angle between the line of sight and the equatorial plane of the accretion disc.  If  the observer is  pole-on for the disc around Sgr A*, the high-velocity jet should be observed because the jet in our model is considerably collimated along the rotational axis.
Conversely, if the observer is edge-on, it may be difficult to detect the powerful jets.
However, even in the edge-on case, the observer could detect considerable emissions  without visible jet component  because the model shows considerable  emission from the outer radial boundary which amounts to 40$\%$ of the total luminosity.

\section{Summary and Discussion}

This paper is the first time that variable properties of low angular momentum magnetized accretion flows around Sgr A* have been studied by solving a relativistic radiation MHD set of equations. We explored the details of radiation processes in such flows in order to explain the observational signatures. Our results can be summarized as follows:

(1) Similarly to the observed phenomena around Sgr A* by Chandra, Swift, and XMM-Newton, the luminosity in the present models varies by order of unity around the average $L \sim 3 \times 10^{34}$ erg $s^{-1}$ and the long-term flares occur approximately every five and ten days on varying plasma beta  $\beta_{\rm out}$ and specific angular momentum $\lambda$.  Almost half of the gas falls onto the black hole while the remaining half  leaves as outflows.

(2) In presence of MHD turbulence, the accretion flow becomes quite asymmetric to the equator. It leads to the intermittent outflow near the horizon, the expanding inner shock, and the quasi-periodically oscillating outer standing shock. During the evolution, both the shocks and the accreting gas interact with each other and lead to complicated features of the luminosity.

(3) The flaring of synchrotron radiation lags behind that of bremsstrahlung radiation by 1-2 hours which is qualitatively compatible with the observed time lags of 1 -- 5 hours in several simultaneous radio and X-ray observations. Both flares are caused by a strong influence of the same thermal processes of shock heating but different radiation processes in the different locations. The time lags can be well explained as the transit time of the acoustic waves between two different regions of the core region of  3 $R_{\rm g}$ size mostly corresponding to the synchrotron emission and the outer region  inside the expanding inner shock which yields the bremsstrahlung emission.

(4) Based on the averaged distribution of radiation, the radiation is anisotropically distributed along the rotational axis and several times stronger in a narrow region of vertical direction than that isotropically distributed in the radial direction. However, the luminosity emitted from the outer z-boundary surface is comparable to
that emitted from the outer R-boundary surface.

(5)  The strong magnetic pressure gradient force leads intermittently to a high-velocity jet with
 $\sim 0.6c$ at the outer z-boundary surface in a narrow funnel region of angle $\sim$ 15 degrees.
  The averaged mass-outflow rate of the jet is significant at roughly 10 percent of the total mass outflow rate. When the oscillating shock expands remotely, the maximum outflow
 rate is obtained but the high-velocity jet appears just after the oscillating shock undergoes maximum contraction.
 The observational detection of the high-velocity jet may be a key if the observer is pole-on or edge-on 
 for the disc around Sgr A*.

The framework of the present model consists of the oscillating outer shock driven by the MRI  and the expanding inner shock in the turbulent flow. 
The shock oscillating model for the long-term flare of Sgr A* explains well some observations 
of the flaring rate and the delay time between radio and X-ray emissions.
However, we notice that Ponti et al. (2017) find a very bright
flare from Sgr A*, which is by more than two orders of magnitude higher than the bremsstrahlung emission in usual flares, starts in
NIR and then an X-ray flare follows after $\sim 10^3$ s, conversely from the time-lag relation  in our model.
They confirm the origin of the very bright flare as the synchrotron nature instead of the bremsstrahlung emission  and suggest another scenario for the bright flare that the electron producing the synchrotron radiation is accelerated by any process tapping energy from the magnetic field such as magnetic reconnection. The scenario may be replaced as a magneto-hydrodynamical model for the episodic mass ejection by Yuan et al. (2009) and Li, Yuan \& Wang (2017).
The low angular momentum of the accretion flow in this paper is considered to be reasonable as far as the inner region of $ R \le 200$ is concerned.
Even if  we consider the high angular momentum flow such as ADAF flows in a far distant region,
the high angular momentum of the flow would be low as $\lambda \sim 1$ near the event horizon, as far as non-rotating black holes are considered (see Nakamura et al. 1996; Manmoto, Mineshige \& Kusunose 1997).
The flow parameters for standing shock formation are constrained to some extent and must be sought from 2D numerical simulations.  However, even if the flow parameters vary a little  from the parameter space responsible for the standing shock, the shock-like behavior of the flow is maintained under the appropriate magnetic field strength.
Proga \& Begelman (2003b) find the time variability of low angular momentum magnetized flows which can account for some of Sgr A* variability.
The assumption of the two-temperature model used in this paper is too simple. If we treat the exact two-temperature
 model which includes radiation processes such as  the energy heating from ions to electrons by Coulomb collision, the bremsstrahlung cooling,  and synchrotron cooling, we can get more realistic temperatures of ion and electron, reproduce time-dependent spectra of Sgr A*, and compare them with the observations. These are subjects of further research in the future.

\section*{Acknowledgments}
CBS is supported by the National Natural Science Foundation of China under grant no. 12073021. The numerical computations were conducted on the Yunnan University Astronomy Supercomputer.

\section*{Data Availability}
The numerically created data that support the findings of this study are available from the corresponding authors upon reasonable request.

\label{lastpage}


\begin{thebibliography}{99}

\bibitem[\protect\citeauthoryear{Aktar、Das \& Nandi }{2015}]{key-1}
 Aktar R., Das S., Nandi A., 2015, MNRAS, 453, 3414
 
 \bibitem[\protect\citeauthoryear{Aktar et al. }{2017}]{key-1-0}
 Aktar R., Das S., Nandi A., Sreehari H., 2017, MNRAS, 471, 4806

\bibitem[\protect\citeauthoryear{Andr\'{e}s et al.}{2022}]{key-1-1}
 Andr\'{e}s A. et al., 2022, MNRAS, 510, 2851

\bibitem[\protect\citeauthoryear{ Balbus \& Hawley }{1991}]{key-2}
 Balbus S. A., Hawley J. F.,  1991, ApJ, 376, 214

\bibitem[\protect\citeauthoryear{Ball et al.} {2016}]{key-4}
Ball D., \"Ozel F., Psaltis D.,  Chan C.-K., 2016, ApJ, 826,77

\bibitem[\protect\citeauthoryear{Becker、Das \& Le}{2011}]{key-5} 
Becker P.~A., Das S., Le T., 2011, ApJ, 743, 47

\bibitem[\protect\citeauthoryear {Bondi } {1952}]{key-6}
Bondi H., 1952, MNRAS, 112, 195

\bibitem[\protect\citeauthoryear {Capellupo et al.} {2017}]{key-6-1}
Capellupo D. M. et al., 2017, ApJ, 845, 35

\bibitem[\protect\citeauthoryear{Chakrabarti }{1989}]{key-7}
 Chakrabarti S. K.,  1989, ApJ, 347, 365

\bibitem[\protect\citeauthoryear{Chakrabarti} {1996}]{key-8}
 Chakrabarti S. K.,  1996, ApJ, 464, 664

\bibitem[\protect\citeauthoryear{Chakrabarti, Acharyya \& Molteni}{2004}]{key-9}
 Chakrabarti S. K., Acharyya K.,  Molteni D., 2004, A\&A, 421, 1

\bibitem[\protect\citeauthoryear{Chakrabarti \& Das} {2004}]{key-9-1}
 Chakrabarti S. K.,  Das S., 2004,  MNRAS, 349, 649

\bibitem[\protect\citeauthoryear{Chan et al. } {2009}]{key-10}
Chan C.-K., Liu S., Fryer C. L., Psaltis D., \"{O}zel F., Rockefeller G.,  Melia F., 2009, ApJ., 701, 521

\bibitem[\protect\citeauthoryear {Czerny \& Mo\'{s}cibrodzka} {2008}]{key-12}
Czerny B.,  Mo\'{s}cibrodzka M., 2008, J. Phys. Conf. Ser.,131, 012001

\bibitem[\protect\citeauthoryear {Das, Becker \& Le.}{2009}]{key-13} 
Das S., Becker P. A.,  Le T., 2009, ApJ, 702, 649


\bibitem[\protect\citeauthoryear{Degenaar et al.} {2013}]{key-15}
Degenaar N., Miller J. M., Kennea J., Gehrels N., Reynolds  M. T.,  Wijnands  R.,  2013, ApJ, 769, 155

\bibitem[\protect\citeauthoryear{Dexter, Agol \& Fragile} {2009}]{key-18}
Dexter J., Agol E.,  Fragile P. C., 2009, ApJ., 703, L142

\bibitem[\protect\citeauthoryear{Dodds-Eden et al.} {2010}]{key-19}
Dodds-Eden K., Sharma P., Quataert E., Genzel R., Gillessen S., Eisenhauer F.,  Porquet D., 2010, ApJ., 725, 450

\bibitem[\protect\citeauthoryear{Eckart et al. }{2006}]{key-20}
Eckart A., Sch\"{o}del R., Meyer L., Trippe S., Ott T.,  Genzel R., 
2006, A\&A, 455, 1

\bibitem[\protect\citeauthoryear{Esin et al. }{1996}]{key-20-1}
Esin A. A., Narayan R., Ostriker E., Yi I., 1996, ApJ, 465, 312

\bibitem[\protect\citeauthoryear{Fukue }{1987}]{key-20-2}
Fukue J., 1987, PASJ, 39, 309




\bibitem[\protect\citeauthoryear{Genzel, Eisenhauer \& Gillessen } {2010}]{key-21}
Genzel R., Eisenhauer F.,  Gillessen S., 2010, Rev. Mod. Phys., 82, 3121

\bibitem[\protect\citeauthoryear{Genzel et al. } {2003}]{key-22}
Genzel R., Sch\"{o}del R., Ott T., Eckart A., Alexander T., 
Lacombe F., Rouan D.,  Aschenbach B., 2003, Nature, 425, 934

\bibitem[\protect\citeauthoryear{Ghez et al. } {2004}]{key-23}
Ghez A. M. et al., 2004, ApJ., 601, L159

\bibitem[\protect\citeauthoryear{Giri et al.}{2010}]{key-23-1}
Giri K., Chakrabarti S. K., Samanta M. M., Ryu D., 2010, MNRAS, 403, 516

\bibitem[\protect\citeauthoryear{Hawley \& Balbus } {1991}]{key-24}
  Hawley J. F.,  Balbus S. A., 1991, ApJ, 376, 223

\bibitem[\protect\citeauthoryear{Igumenshchev \& Abramowicz } {1999}]{key-25}
 Igumenshchev  I. V.,  Abramowicz M. A., 1999, MNRAS, 303, 309 

\bibitem[\protect\citeauthoryear{Igumenshchev \& Abramowicz} {2000}]{key-26}
 Igumenshchev I. V.,  Abramowicz M. A., 2000, ApJS, 130, 463

\bibitem[\protect\citeauthoryear{Igumenshchev, Narayan \& Abramowicz } {2003}]{key-27}
 Igumenshchev I. V., Narayan R.,  Abramowicz M. A.,  2003, ApJ,  592, 1042

\bibitem[\protect\citeauthoryear{Kumar \& Chattopadhyay} {2013}]{key-27-1}
 Kumar R., Chattopadhyay I.,  2013, MNRAS, 430, 386

\bibitem[\protect\citeauthoryear{Lanzafame, Molteni \& Chakrabarti}{1998}]{key-27-2}
Lanzafame G., Molteni D., Chakrabarti S. K., 1998, MNRAS., 299, 799

\bibitem[\protect\citeauthoryear{Li, Ostriker \& Sunyaev} {2013}]{key-29}
Li J., Ostriker J.,  Sunyaev R., 2013, ApJ., 767, 105

\bibitem[\protect\citeauthoryear{Li, Yuan \& Wang } {2017}]{key-30}
Li Y.-P., Yuan F.,  Wang Q. D., 2017, MNRAS, 468, 2552

\bibitem[\protect\citeauthoryear{Loeb} {2004}]{key-30-1}
Loeb A., 2004,  MNRAS, 350, 725

\bibitem[\protect\citeauthoryear{Machida, Hayashi \& Matsumoto} {2000}]{key-31}
 Machida M., Hayashi M. R.,  Matsumoto R.,  2000, ApJ, 532, L67

\bibitem[\protect\citeauthoryear{Machida, Matsumoto \& Mineshige} {2001}]{key-32}
 Machida M., Matsumoto R.,  Mineshige S., 2001, PASJ, 53, L1

\bibitem[\protect\citeauthoryear{Manmoto, Mineshige \& Kusunose} {1997}]{key-32-1}
 Manmoto T., Mineshige S., Kusunose M., 1997 ApJ, 489, 791
 
 \bibitem[\protect\citeauthoryear{Melon Fuksman \& Mignone } {2019}]{key-32-2}
Melon Fuksman J. D., Mignone A., 2019, ApJS, 242, 20

\bibitem[\protect\citeauthoryear{Meyer et al. } {2006b}]{key-34}
Meyer L.,  Eckart A., Sch\"{o}del R.,  Duschl W. J., Mu\v{z}i\'{c} K., Dov\v{c}iak M.,  Karas V., 2006a, A\&A, 460, 15

\bibitem[\protect\citeauthoryear{Meyer et al. } {2006a}]{key-33}
Meyer L., Sch\"{o}del R., Eckart A., Karas V., Dov\v{c}iak M.,  Duschl W. J.,
2006b, A\&A, 458, L25

\bibitem[\protect\citeauthoryear{Mignone et al. } {2007}]{key-35}
Mignone A., Bodo G., Massaglia S., Matsakos T., Tesileanu O., Zanni C.,  Ferrari A., 2007, ApJS, 170, 228

\bibitem[\protect\citeauthoryear{Molteni, Lanzafame \& Chakrabarti}{1994}]{key-35-1}
Molteni D., Lanzafame G.,  Chakrabarti S. K., 1994, ApJ., 425, 161

\bibitem[\protect\citeauthoryear{Molteni, Sponholz \& Chakrabarti}{1996}]{key-35-2}
Molteni D., Sponholz H., Chakrabarti S. K., 1996, ApJ., 457, 805

\bibitem[\protect\citeauthoryear{Molteni, Ryu \& Chakrabarti}{1996}]{key-35-3}
Molteni D., Ryu D., Chakrabarti S. K., 1996, ApJ., 470, 460

\bibitem[\protect\citeauthoryear{Mondal \& Chakrabarti}{2006}]{key-35-4}
 Mondal S., Chakrabarti S.,  2006, MNRAS, 371, 1418


\bibitem[\protect\citeauthoryear{Mo\'{s}cibrodzka, Das  Czerny} {2006}]{key-36}
Mo\'{s}cibrodzka M., Das T.K.,  Czerny B., 2006, MNRAS, 370, 219

\bibitem[\protect\citeauthoryear{Nakamura et al.} {1996}]{key-36-1}
 Nakamura K. E., Matsumoto R., Kusunose M.,  Kato S., 1996, PASJ, 48, 761

\bibitem[\protect\citeauthoryear{Narayan, Igumenshchev \& Abramowicz} {2003}]{key-38}
 Narayan R., Igumenshchev I. V.,  Abramowicz M. A., 2003, PASJ, 55, L69

\bibitem[\protect\citeauthoryear{Narayan \& McClintock } {2008}]{key-39}
 Narayan R.,  McClintock J. E.,  2008, New Astron. Rev., 51, 733

\bibitem[\protect\citeauthoryear{Narayan et al. } {2012}]{key-40}
 Narayan R., Sadowski A., Penna R. F.,  Kulkarni A. K., 2012, MNRAS, 426, 3241
 
 \bibitem[\protect\citeauthoryear{Narayan \& Yi } {1994}]{key-41}
 Narayan R.,  Yi I.,  1994, ApJ, 428, L13

\bibitem[\protect\citeauthoryear{Narayan \& Yi } {1995}]{key-42}
 Narayan R.,  Yi I., 1995, ApJ, 452, 710

\bibitem[\protect\citeauthoryear{Neilsen et al. } {2013}]{key-43}
 Neilsen J. et al., 2013, ApJ, 774, 42

\bibitem[\protect\citeauthoryear{Neilsen et al. } {2015}]{key-44}
 Neilsen J. et al., 2015, ApJ, 799, 199

 \bibitem[\protect\citeauthoryear{Okuda }{2014}]{key-45}
 Okuda T.,  2014, MNRAS, 441, 2354
 
 \bibitem[\protect\citeauthoryear{Okuda \& Das } {2015}]{key-45-1}
 Okuda T.,   Das S., 2015, MNRAS, 453, 147
 

\bibitem[\protect\citeauthoryear{Okuda \& Molteni }{2012}]{key-45-2}
 Okuda T.,  Molteni D.,  2012, MNRAS, 425, 2413

\bibitem[\protect\citeauthoryear{Okuda et al.} {2019}]{key-46}
 Okuda T., Singh C. B., Das S., Aktar R., Nandi A., deGouveia Dal Pino E. M., 2019, PASJ, 71, 49

\bibitem[\protect\citeauthoryear{Paczy\'{n}sky \&  Wiita } {1980}]{key-48}
  Paczy\'{n}sky B.,  Wiita P. J.,  1980, A\&A, 88, 23

 \bibitem[\protect\citeauthoryear{Ponti et al. } {2015}]{key-49}
 Ponti G. et al., 2015, MNRAS, 454, 1525

 \bibitem[\protect\citeauthoryear{Ponti et al. } {2017}]{key-50}
 Ponti G. et al., 2017, MNRAS, 468, 2447

\bibitem[\protect\citeauthoryear{Proga \& Begelman }{2003a}]{key-51-1}
 Proga D.,  Begelman M. C. 2003a, ApJ, 582, 69

\bibitem[\protect\citeauthoryear{Proga \& Begelman } {2003b}]{key-51-2}
 Proga D.,  Begelman M. C., 2003b, ApJ, 592, 767

\bibitem[\protect\citeauthoryear{Rauchi et al. } {2016}]{key-51-3}
 Rauch C., Ros E., Krichbaum T. P., Eckart A., Zensus J. A., Shahzamanian B.,
 Mu\v{z}i\'{c}  K.,  2016, A\&A, 587, A37
 
\bibitem[\protect\citeauthoryear{Ressler et al. } {2017}]{key-52}
 Ressler S. M., Tchekhovskoy A., Quataert E.,  Gammie C. F., 2017, MNRAS, 467, 3604

\bibitem[\protect\citeauthoryear{Roberts et al. } {2017}]{key-53}
 Roberts S. R., Jiang Y.-F., Wang Q. D.,  Ostriker J. P., 2017, MNRAS, 466, 1477
 
\bibitem[\protect\citeauthoryear{Sarkar \& Das } {2016}]{key-54} 
Sarkar B.,  Das S., 2016, MNRAS, 461, 190

\bibitem[\protect\citeauthoryear{Shakura \& Sunyaev } {1973}]{key-55}
 Shakura N. I.,  Sunyaev R. A., 1973, A\&A, 24, 337

\bibitem[\protect\citeauthoryear{Singh \& Chakrabarti } {2011}]{key-56}
 Singh C. B.,  Chakrabarti S. K., 2011, MNRAS, 410, 2414

\bibitem[\protect\citeauthoryear{Singh, Okuda \& Aktar } {2021}]{key-56-1}
 Singh C. B., Okuda T., Aktar R., 2021, RAA, 21, 134

\bibitem[\protect\citeauthoryear{Stepney \& Guilbert } {1983}]{key-56-2}
 Stepney S.,  Guilbert P. W., 1983, MNRAS, 204, 1269

\bibitem[\protect\citeauthoryear{Stone \& Pringle } {2001}]{key-57}
 Stone J. M.,  Pringle J. E., 2001, MNRAS, 322, 461

\bibitem[\protect\citeauthoryear{Stone, Pringle \& Begelman } {1999}]{key-58}
 Stone J. M., Pringle J. E.,  Begelman M. C.,  1999, MNRAS, 310, 1002

\bibitem[\protect\citeauthoryear{Trippe et al. } {2007}]{key-59}
 Trippe S., Paumard T., Ott T., Gillessen S., Eisenhauer F., Martins F., Genzel R., 2007, MNRAS, 375, 764


\bibitem[\protect\citeauthoryear{Wang et al. } {2013}]{key-59-1}
 Wang Q. D. et al., 2013, Science, 341, 981

\bibitem[\protect\citeauthoryear{Witzel et al. } {2021}]{key-59-2}
 Witzel G. et al., 2021, ApJ, 917, 73

\bibitem[\protect\citeauthoryear{Yuan} {(2011}]{key-60}
Yuan F., 2011, in Morris M. R., Wang  Q. D., Yuan F., eds, 
ASP Conf. Ser. Vol. 439,
The Galactic Center: A Window to the Nuclear Environment of Disk 
 Galaxies. Astron. Soc. Pac., San Francisco, p. 346

\bibitem[\protect\citeauthoryear{Yuan, Bu \& Wu } {2012}]{key-61}
Yuan F., Bu D.,  Wu M.,  2012, ApJ, 761, 130

\bibitem[\protect\citeauthoryear{Yuan et al. } {2015}]{key-62}
Yuan F., Gan Z., Narayan R., Sadowski A., Bu D.,  Bai X.-N., 2015, ApJ, 804,101

\bibitem[\protect\citeauthoryear{Yuan et al. } {2009}]{key-63}
Yuan F., Lin J., Wu K.,  Ho L. C., 2009, MNRAS, 395, 2183

\bibitem[\protect\citeauthoryear{Yuan \& Narayan } {2014}]{key-64}
Yuan F.,  Narayan R.,  2014, ARA\&A, 52, 529

\bibitem[\protect\citeauthoryear{Yuan, Quataert \& Narayan} {2003}]{key-65}
Yuan F., Quataert E.,  Narayan R.,  2003, ApJ, 598, 301

\bibitem[\protect\citeauthoryear{Yuan, Quataert \& Narayan } {2004}]{key-66}
Yuan F., Quataert E.,  Narayan R.,  2004, ApJ, 606, 894

\bibitem[\protect\citeauthoryear{Yuan, Wu \& Bu } {2012}]{key-67}
Yuan F., Wu M.,  Bu D.,  2012, ApJ, 761, 129

\bibitem[\protect\citeauthoryear{Yusef-Zadeh et al. } {2006}]{key-68}
Yusef-Zadeh F. et al., 2006, ApJ, 644, 198


\bibitem[\protect\citeauthoryear{Yusef-Zadeh et al. } {2008}]{key-69}
Yusef-Zadeh F., Wardle M., Heinke C., Dowell C. D., Roberts D., Baganoff F. K., Cotton W., 2008, ApJ, 682, 361

\bibitem[\protect\citeauthoryear{Yusef-Zadeh et al. } {2009}]{key-70}
Yusef-Zadeh F. et al., 2009, ApJ, 706, 348

\bibitem[\protect\citeauthoryear{Yusef-Zadeh et al. } {2011}]{key-71}
Yusef-Zadeh F., Wardle M., Miller-Jones J. C. A.,  Roberts D.A., 
Grosso N.,  Porquet D.,  2011, ApJ, 729, 44

\end{thebibliography}
\end{document}